\documentclass[%
 reprint,
 superscriptaddress,
 amsmath,amssymb,
 aps,
 floatfix,
 mathtools,
]{revtex4-2}

\usepackage{eqnarray}\usepackage{amsthm}
\usepackage{fancyhdr}
\usepackage{subfigure}\usepackage{epsfig}
\usepackage{epsfig}
\usepackage{bbm}
\usepackage{subfigure}
\usepackage{bm}
\usepackage{float}
\usepackage{xcolor}

\usepackage{graphicx}% Include figure files
\usepackage{dcolumn}% Align table columns on decimal point
\usepackage{bm,braket}% bold math
\newcommand{\I}{\boldsymbol{1}}
\begin{document}

\preprint{}

\title{Active chiral molecules in activity gradients}% Force line breaks with \\

\author{Pietro Luigi Muzzeddu}
\affiliation{SISSA--International School for Advanced Studies, I-34136 Trieste, Italy}%
\author{Hidde Derk Vuijk}
\affiliation{Leibniz-Institut f\"ur Polymerforschung Dresden, Institut Theory der Polymere, 01069 Dresden, Germany}
\author{Hartmut L\"owen}
\affiliation{Institut f\"ur Theoretische Physik II : Weiche Materie, Heinrich-Heine-Universit\"at D\"usseldorf, 40225 D\"usseldorf, Deutschland }
\author{Jens-Uwe Sommer}
\affiliation{Leibniz-Institut f\"ur Polymerforschung Dresden, Institut Theory der Polymere, 01069 Dresden, Germany}
\affiliation{Technische Universit\"at Dresden, Institut f\"ur Theoretische Physik, 01069 Dresden, Germany}
\author{Abhinav Sharma}
\affiliation{Leibniz-Institut f\"ur Polymerforschung Dresden, Institut Theory der Polymere, 01069 Dresden, Germany}
\affiliation{Technische Universit\"at Dresden, Institut f\"ur Theoretische Physik, 01069 Dresden, Germany}

%\affiliation{
%\affiliation{%


\date{\today}% It is always \today, today,

\begin{abstract}
While the behavior of active colloidal molecules is well studied by now for a constant activity, the effect of activity gradients is much less understood. Here we explore one of the simplest molecules in activity gradients, namely active chiral dimers composed of two particles with opposite active torques of the same magnitude. We show analytically that with increasing torque, the dimer switches its behavior from antichemotactic to chemotactic. The origin of the emergent chemotaxis is the cooperative exploration of activity gradient by the two particles. While one of the particles moves into higher activity regions, the other moves towards lower activity region resulting in a net bias in the direction of higher activity. We do a comparative study of chiral active particles with charged Brownian particles under magnetic field and show that despite the fundamental similarity in terms of their odd-diffusive behavior, their dynamics and chemotactic behavior are generally not equivalent. We demonstrate this explicitly in a dimer composed of oppositely charged active particles, which remains antichemotactic for any magnetic field.
\end{abstract}

\maketitle

%\tableofcontents
\section{\label{sec:level1} Introduction}
Living matter at the micron scale is able to perform a wide variety of complex motions and behaviors, see e.g. \cite{ahamed2021capturing,costa2019adaptive}.
This requires sensing chemical and structural properties of the environment, processing this information by complex biochemical networks, and adapting the behavior accordingly~\cite{wadhams2004making,tu2013quantitative,sengupta2021principles}. For example, the bacterium \emph{E. coli} measures the local nutrient concentration while it swims, and compares that with the concentration in the past. If the current concentration is getting lower (higher), it increases (decreases) its tumble rate. This strategy results in chemotaxis, i.e., preferential accumulation in regions where the nutrient concentration is high \cite{berg2008coli,cates2012diffusive}.

Active particles are regarded as the simplest models for motility in living systems. While they share the essential features, namely self-propulsion and persistence, with the bacteria, their response to local fuel concentration is rather simple; merely adjust the speed in proportion to the local fuel concentration \cite{howse2007selfmotile, gao2014catalytic, jee2018catalytic, stenhammar2016lightinduced, caprini2022dynamics}. As a consequence, active particles subjected to fuel gradients, preferentially accumulate in regions where the fuel concentration is low \cite{schnitzer1993theory, sharma2017brownian}. Nevertheless, the ability to steer active particles towards correct target zone -- artificial chemotaxis -- remains a highly sought property of synthetic active matter. While this has been demonstrated experimentally via elaborate feedback mechanisms \citep{mano2017optimal,qian2013harnessing,massana2022optical}, where an external stimulus is applied to the particle as a function of its state (position and orientation), the continuous monitoring of the particle's state might not always be possible. Therefore, an autonomous approach where 
some form of feedback emerges spontaneously is much more desirable, and can correctly steer the active particle towards the target without requiring external stimuli. 

Recently, we showed that chemotaxis can emerge in a system of cargo carrying active particles~\cite{vuijk2021chemotaxis}. While active particles with a light cargo accumulate in regions of low activity, a crossover occurs with increasing cargo such that the active-passive complex accumulates in regions of high activity. For a heavy cargo which moves much more slowly than the active particle, the active particle \emph{performs} a local integration of the activity profile in the neighborhood of the attached cargo resulting in a net force towards the regions of higher activity giving rise to chemotactic behavior. Active particles connected in a chain to form polymers, have recently received much attention  \citep{winkler2017active,mousavi2019active,winkler2020physics,martin-gomez2019active,eisenstecken2022path,kaiser2015how}. We showed that active polymers are qualitatively similar to a single active particle coupled to a passive cargo  \cite{vuijk2021chemotaxis}. While short chains are antichemotactic, chains with $4$ or more particles exhibit chemotactic behavior.

\begin{figure}[t]
\includegraphics[scale=0.28]{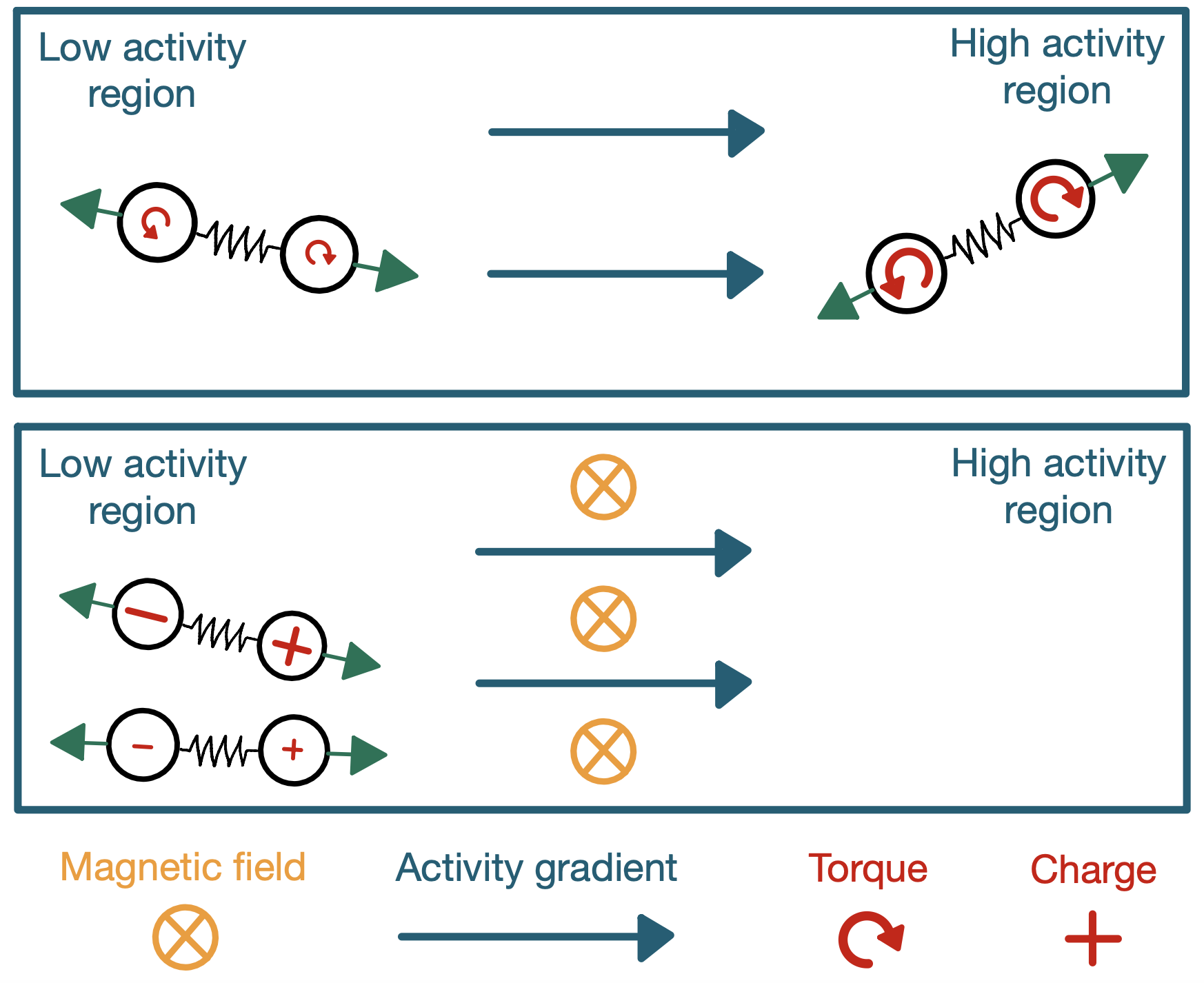}
\caption{Dimers in activity gradients. In the upper panel, both particles experience homogeneous torques in the direction perpendicular to the plane of motion giving rise to their chiral motion (curved arrows).  When the active torque is small the dimer accumulates in the region of low activity (antichemotactic). Beyond a threshold active torque, the dimer switches its tactic behavior, i.e., it accumulates in the region of high activity. The lower panel shows a dimer composed of oppositely charged particles. While the particles experience an active force, there is no active torque acting on them. Instead, the dimer is subjected to a homogeneous external magnetic field. The charged active dimer always shows antichemotactic behavior, i.e., unlike its torque-driven counterpart, there is no magnetic field governed crossover in its tactic behavior.}
\label{schematic_dimers}
\end{figure}

In this paper, we extend the idea of emergent chemotaxis to colloidal molecules made of active chiral particles (ACPs). Specifically, we consider a dimer composed of two active chiral particles driven by opposite torques in an activity gradient (see schematic in Figure \ref{schematic_dimers}). With increasing torque, the active chiral dimer switches its behavior from antichemotactic to chemotactic and accumulates in regions of high activity. While the behaviour of active colloidal molecules is well understood for a constant activity~\cite{ebbens2010selfassembled,lowen2018active}, the effect of activity gradients is much less explored. Very recently, we studied how a rigid dimer composed of two active particles with orientations fixed with respect to the connecting bond behaves in activity gradients \cite{vuijk2022active}. In contrast, in an active chiral dimer, the orientations of the two particles evolve freely due to the thermal and chiral torques.

Active chiral particles exhibit odd-diffusive motion on time scales greater than the persistence time \cite{hargus2021odd}. The diffusion tensor that characterizes the overdamped motion contains both a symmetric and an antisymmetric part. Both passive and self-propelled charged Brownian particles also perform odd-diffusive motion under the effect of the Lorentz force \cite{vuijk2020lorentz,abdoli2020nondiffusive,abdoli2020stationary,abdoli2020correlations,abdoli2021brownian}. While active chiral particles rotate due to the microscopic active torque, in the case of charged particles under magnetic field a certain handedness is introduced by the Lorentz force. On a coarse grained level, where one integrates out the orientational degree of freedom, the two systems present many similarities. One can even map the two model systems for homogeneous magnetic field, activity and torque. However, as expected the mapping does not hold in general. We show this explicitly analyzing their chemotactic behavior: a dimer composed of oppositely charged active particles always accumulates in the regions of low activity independently of the applied magnetic field.\\
The paper is organized as follows. In section~\ref{ACD}, we present our model of a dimer composed of two active chiral particles driven by opposite torques of the same magnitude. We derive a coarse grained Fokker-Planck equation from which we obtain an analytical prediction for the density distribution of the dimer. In section~\ref{ODD}, inspired by the similarity between an ACP and a charged self propelled particle subjected to Lorentz force in terms of their odd diffusive behaviour, we construct a dimer of these active charged particles and show that unlike a dimer of ACPs with opposite torques, there is no magnetic field governed crossover in the tactic behavior. Finally, in section~\ref{discussion_conclusions}, we discuss possible experimental realisations and present a brief outlook for the future work.
\section{Dimer of active chiral particles}
\label{ACD}
We consider a two dimensional system of a dimer composed of two active chiral particles interacting via an attractive potential $U(\bm{r})$ (e.g. a spring or a rigid rod). In addition to (thermal) rotational diffusion, both particles experience homogeneous torques in the direction perpendicular to the plane of motion, giving rise to their chiral motion. We restrict our analysis to opposite torques $(\omega,-\omega)$ on the two active particles. The overdamped dynamics of the system are governed by the following Langevin equations~\cite{bechinger2016active}:
\begin{equation}
\begin{split}
       \frac{d \bm{r}_1}{dt}&=-\frac{1}{\gamma} \nabla_1 U(\bm{r}_1-\bm{r}_2) + v_a(\bm{r}_1) \bm{p}_1 + \sqrt{2 D_T} \bm{\xi}_1(t),\\
    \frac{d \bm{r}_2}{dt}&=+ \frac{1}{\gamma} \nabla_1 U(\bm{r}_1-\bm{r}_2) + v_a(\bm{r}_2) \bm{p}_2 + \sqrt{2 D_T} \bm{\xi}_2(t),\\
    \frac{d \theta_1}{dt} &=+ \omega + \sqrt{2 D_R} \eta_1(t),\\
    \frac{d \theta_2}{dt} &= -\omega + \sqrt{2 D_R} \eta_2(t),
    \end{split}
    \label{Langevin}
\end{equation}
where $\bm{p}_i=(\cos(\theta_i),\sin(\theta_i))$ are the orientation vectors and $\bm{\xi}_1(t),\bm{\xi}_2(t),\eta_1(t),\eta_2(t)$ are independent random Gaussian vectors with the following statistical properties:
\begin{equation}
    \begin{split}
    &\braket{\bm{\xi}_{1}(t)}=\braket{\bm{\xi}_{2}(t)}=\braket{\bm{\eta}(t)}=0,\\
    &\braket{\bm{\xi}_{1}(t)\bm{\xi}^T_{1}(s)}=\braket{\bm{\xi}_{2}(t)\bm{\xi}^T_{2}(s)}=\bm{1}\delta(t-s),\\
    &\braket{\eta_1(t)\eta_1(s)}=\braket{\eta_2(t)\eta_2(s)}=\delta(t-s).\\
    \end{split}
\end{equation}
%  \sn{Hey Pier, I suggest to use vector notion for noise properties, i.e., $\braket{\bm{\eta}(t)}=0$ and $\braket{\bm{\eta}^T(t)\bm{\eta}(t')}=\I\delta(t-t')$}
 
 \begin{figure*}[t]
\includegraphics[scale=0.43]{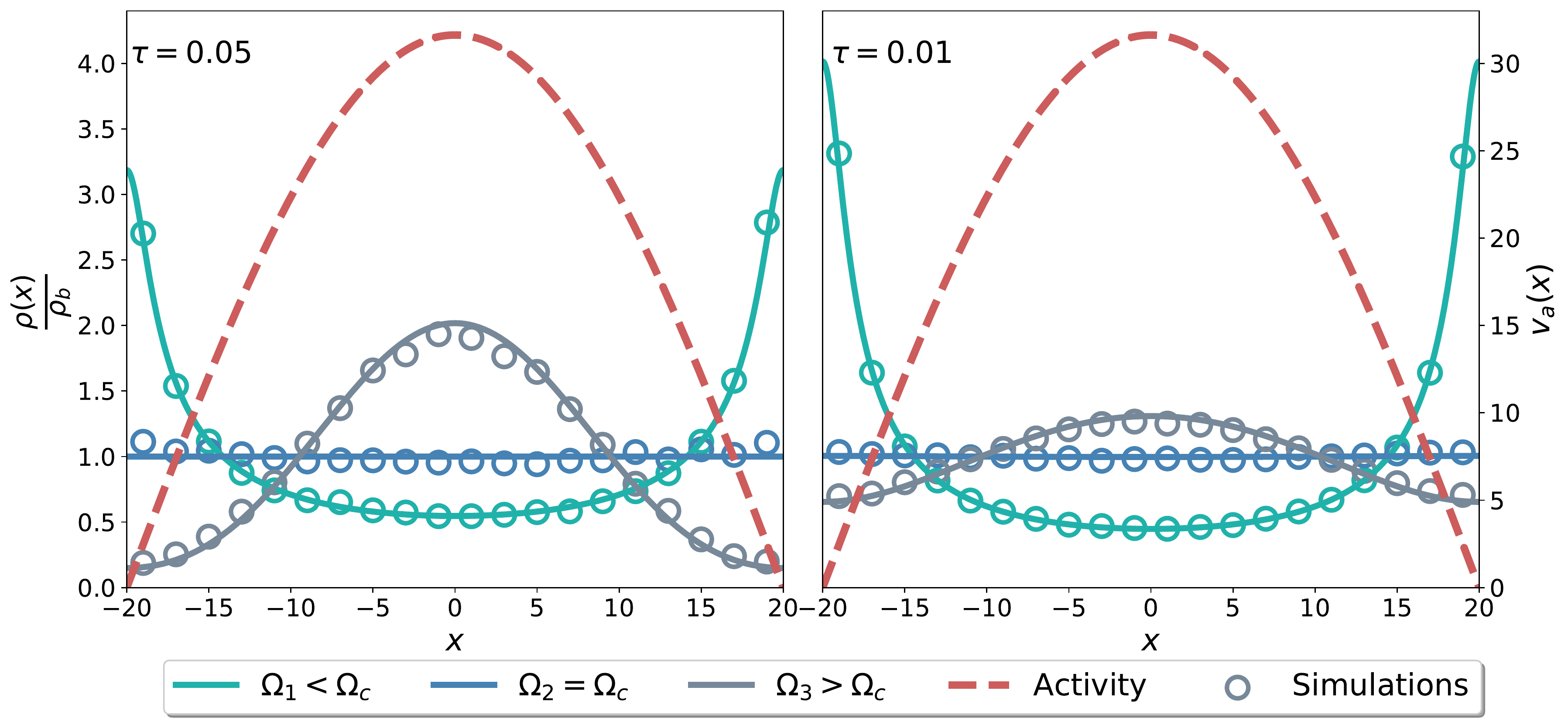}
\caption{Steady state density of dimers made by active chiral particles with opposite torque. All lengths and times are measured in units of $l=\sqrt{D_T/D_R}$ and $\tau_r=1/D_R$, respectively. Both analytical (continuous line) and numerical (symbols) results have been obtained adopting  periodic boundary conditions. The dashed red curve represents the activity profile, which in this case is homogeneous along $y$-direction and varying according to $v_a(x)=10\sqrt{10}\sin(\frac{\pi}{2L}x+\frac{\pi}{2})$ (in units of $\sqrt{D_T D_R}$) along $x$-direction, with $2L$ the elementary cell size. Since the system is invariant to translations along the $y$-direction, we report here the marginal distribution $\rho(x)$ divided by the bulk density. The two figures are characterized by different spring relaxation time scales $\tau$. In particular, for higher stiffness, the density peak in the chemotactic regime is broader and the dimer is less localized. Simulations have been carried out with the following parameters: $\Omega_1=2<\Omega_c$, $\Omega_2=4.58257\approx \Omega_c$, $\Omega_3=10>\Omega_c$, $\tau=0.05$ on the left panel, and $\Omega_1=2<\Omega_c$, $\Omega_2=10.05\approx \Omega_c$, $\Omega_3=15>\Omega_c$, $\tau=0.01$ on the right panel. }
\label{densityprediction}
\end{figure*}

The two ACPs composing the dimer have the same friction coefficient $\gamma$, translational diffusion coefficient $D_T=\frac{k_B T}{\gamma}$ and rotational diffusion coefficient $D_R$. In our model, the rotational diffusion coefficient is a free parameter \cite{farage2015effective, caprini2020active, sharma2017escape}. The two ACPs are self propelled along the direction given by their orientation vectors $(\bm{p}_1,\bm{p}_2)$ with an intensity set by the activity function $v_a(\bm{r})$, which we assume being positively correlated with a local fuel concentration. The additional torque acting on each particle produces a systematic rotation of the orientation vector around the fixed $z$-axis. Note that we do not take into account the torque on the two active particles due to the activity gradient because this depends on the specific self-propulsion mechanism
\citep{golestanian2009anomalous,lozano2016phototaxis}.
However, this torque can be included in the analysis presented here.
We also ignore the hydrodynamic interaction between the two particles, and their
effect on the self-propulsion
\citep{popescu2011pulling,popescu2018effective,reigh2015catalytic,reigh2018diffusiophoretically}.

The relative importance of the deterministic and stochastic contributions to the $\bm{p}_i$ (or $\theta_i$) dynamics plays a crucial role in determining the behaviour of our system, as shown in the following. Since we are interested in where dimers accumulate at steady state, we aim to find an expression for the coarse-grained density as a function of the dimer's center of mass only. After rewriting the dynamics in terms of the collective coordinate $\bm{R}=\frac{\bm{r}_1+\bm{r}_2}2$ and the inner coordinate $\bm{r}=\bm{r}_1-\bm{r}_2$, the related Fokker-Planck equation reads as
\begin{widetext}
\begin{equation}
    \begin{split}
        &\partial_t P(\bm{R},\bm{r},\theta_1,\theta_2,t) = - \nabla_{\bm{R}} \cdot \Big[ \frac{1}{2}v_a \Big(\bm{R}+\frac{\bm{r}}{2} \Big)\bm{p}_1 P \Big] - \nabla_{\bm{R}} \cdot \Big[ \frac{1}{2}v_a \Big(\bm{R}-\frac{\bm{r}}{2}\Big)\bm{p}_2 P -\frac{D_T}{2} \nabla_{\bm{R}} P \Big] -\nabla_{\bm{r}} \cdot \Big[ \frac{2}{\gamma} \bm{F}P \Big] + \\ 
        & -\nabla_{\bm{r}} \cdot \Big[ v_a \Big(\bm{R}+\frac{\bm{r}}{2} \Big)\bm{p}_1 P - v_a \Big(\bm{R}-\frac{\bm{r}}{2}\Big)\bm{p}_2 P - 2 D_T \nabla_{\bm{r}} P \Big]
        - \frac{\partial}{\partial \theta_1} \Big[ \omega P - D_R \frac{\partial P}{\partial \theta_1} \Big] - \frac{\partial}{\partial \theta_2} \Big[ -\omega P - D_R \frac{\partial P}{\partial \theta_2} \Big].
    \end{split}
\end{equation}
\end{widetext}
One can now proceed by expanding the joint probability distribution $P(\bm{R},\bm{r},\theta_1,\theta_2,t)$ into the eigenfunctions of the operator $\hat{\mathcal{R}}^2=\frac{\partial^2}{\partial \theta_1^2} + \frac{\partial^2}{\partial \theta_2^2}$. This angular multipole expansion leads to the following expression:
\begin{equation}
    \begin{split}
        &P(\bm{R},\bm{r},\theta_1,\theta_2,t) = \frac{1}{\Omega_2^2} \Big[\phi + \bm{\sigma}_1 \cdot \bm{p}_1 + \bm{\sigma}_2 \cdot \bm{p}_2 + \underline{\underline{\bm{\sigma}}}_{1,2} : \bm{p}_1\bm{p}_2 \\ & + \underline{\underline{\bm{w}}}_1 : \Big( \bm{p}_1\bm{p}_1 - \frac{\bm{1}}{2}\Big) + 
         \underline{\underline{\bm{w}}}_2 : \Big( \bm{p}_2\bm{p}_2 - \frac{\bm{1}}{2}\Big) + ... \Big].
    \end{split}
\end{equation}
A hierarchy of equations for the coefficients $(\phi,\bm{\sigma}_1,\bm{\sigma}_2, \underline{\underline{\bm{\sigma}}}_{1,2},\underline{\underline{\bm{w}}}_1,\underline{\underline{\bm{w}}}_2,...)$ can be obtained by projecting the FPE onto their relative eigenfunctions. We focus on the case where the activity function is slowly varying in space. This small gradient assumption allows us to decouple the equations for the various coefficients in the hierarchy and to find an effective equation for the coarse grained density
\begin{equation}
    \rho(\bm{R})= \int d\bm{r}\, \phi(\bm{R},\bm{r}).
\end{equation}
Details of the coarse graining procedure and the calculation of the steady-state properties are shown in the Supplementary material, Section $1$.
In particular, when the attractive interaction corresponds to a spring with zero rest length, i.e., $U(\bm{r})=\frac{k}{2}\bm{r}^2$, the coarse grained density, $\rho(\bm{R})$, satisfies the following Fokker-Planck equation:
\begin{equation}
    \begin{split}
    &\partial_t \rho(\bm{R})= -\nabla_{\bm{R}} \cdot \Big[ \bm{V}(\bm{R})\rho(\bm{R}) - D(\bm{R}) \nabla_{\bm{R}}\rho(\bm{R}) \Big],\\
    \\
    &D(\bm{R})=\frac{1}{1 + \Omega^2} \frac{v_a^2(\bm{R})}{4D_R} + \frac{D_T}{2},\\
    \\
    &\bm{V}(\bm{R}) = -\frac{1}{2} \epsilon \nabla_{R} D(\bm{R}),\\
    \end{split}
    \label{VandD}
\end{equation}
where $\Omega = \omega/D_R$. The parameter $\epsilon$ is defined in terms of dimensionless parameters $\tau = \gamma D_R/2k$ and $\Omega$ as
% \begin{equation}
%     \epsilon=\frac{\big[D_R+\big( D_R^2-\omega^2\big)\tau]\frac{D_R^2 + \omega^2}{D_R}}{\big[D_R+\big( D_R^2-\omega^2\big)\tau\big]^2 + \big[ \omega\left(1+2D_R \tau \right)\big]^2}.
% \end{equation}
\begin{equation}
    \epsilon=\frac{\left(1+(1-\Omega^2)\tau\right)\left(1+\Omega^2 \right)}{\left(1+(1-\Omega^2)\tau\right)^2 + \left( \Omega\left(1+2 \tau \right)\right)^2}.
\end{equation}
$\Omega$ and $\tau$ express the active torque and the relaxation time of the spring in units of the rotational diffusion time scale, $D_R^{-1}$. Since the drift and diffusion terms are related by a derivative relation, it is possible to find an expression for the steady state density without specifying the form of the activity landscape. By imposing zero flux condition along the direction in which activity varies, one obtains for the density
% \begin{equation}
%     \rho(\mathbf{R}) \propto \Big[ \frac{1}{4}\frac{D_R}{D_R^2 + \omega^2} v_a^2(\mathbf{R}) + \frac{D_T}{2} \Big]^{-\frac{1}{2}\epsilon}.
%     \label{steady_state1}
% \end{equation}
\begin{equation}
    \rho(\mathbf{R}) \propto \Big[1 +  \frac{1}{D_T}\frac{1}{1 + \Omega^2} \frac{v_a^2(\mathbf{R})}{2D_R} \Big]^{-\frac{1}{2}\epsilon}.
    \label{steady_state1}
\end{equation}
\textcolor{blue}{}In absence of activity $(v_a(\bm{R})=0)$ the system reduces to a dimer of passive Brownian particles, so we expect a standard diffusive process and a flat distribution at steady state. Notice that if $v_a(\bm{R})=0$, translational and rotational degrees of freedom are decoupled, and the particle's orientation vector does not play any role in the dynamics. In the case of spatially homogeneous activity, the symmetry of the system imposes again flat steady state distribution. Indeed, the spatial dependence of $\rho(\bm{R})$ in (\ref{steady_state1}) is a consequence of the broken spatial invariance due to the inhomogeneous activity landscape.

For a generic spatially varying $v_a(\bm{R})$, the way activity affects the steady state density profile depends on the sign of $\epsilon$
which in turn is determined by the relative importance of the following two terms: 1+ $\tau$ and $\Omega^2 \tau$. For a fixed spring relaxation time, the competition between rotational diffusion and deterministic active rotation of $\bm{p}_i$ can lead to qualitatively different scenarios. In particular if $1+ \tau > \Omega^2 \tau$, the exponent $\epsilon$ is positive and the density peaks in regions of lower activity. In the limiting case of a vanishing active torque, the exponent $\epsilon \rightarrow 1/(1+\tau)$. Since the activity profile reflects the local fuel concentration, we refer to this behavior as antichemotaxis. On the contrary, if $1+\tau < \Omega^2 \tau$ dimers accumulate in high activity regions. In the limit of $\omega \rightarrow \infty$, the exponent $\epsilon \rightarrow -1/\tau$. We call this phenomenon chemotaxis. The crossover between these two regimes occurs at the critical value:
\begin{equation}
    \Omega_c=\frac{\omega_c}{D_R}=\sqrt{\frac{1+\tau}{\tau}},
    \label{critical_curve}
\end{equation}
at which the steady state is characterized by a homogeneous density distribution.  Figure~\ref{densityprediction} shows the stationary density distribution of dimers for three different values of $\omega$ obtained from simulations of Eq.~\eqref{Langevin}. The theoretical predictions of Eq.~\eqref{steady_state1} are in excellent agreement with the simulations.

% $D_R\big( 1+D_R \tau \big)$ and $\omega^2 \tau$. As anticipated, at fixed $\tau$, the competition between rotational diffusion and deterministic systematic rotation of $\bm{p}_i$ can lead to qualitatively different scenarios. In particular, if $D_R\big( 1+D_R \tau\big) > \omega^2 \tau$ the exponent $\epsilon$ is positive and the density peaks in regions of lower activity. Since the activity profile reflects the local fuel concentration, we refer to this behaviour as antichemotaxis. On the contrary, if $D_R\big( 1+D_R \tau \big) < \omega^2 \tau$ we have accumulation of dimers in high activity regions. We call this phenomenon chemotaxis. The crossover between these two regimes occurs when the torque acting on the dimer components assumes the critical value:
% \begin{equation}
%     \omega_c=\sqrt{D_R\Big( \frac{1}{\tau}+D_R\Big)},
%     \label{critical_curve}
% \end{equation}
% at which the steady state is characterized by a homogeneous density distribution. Figure~\ref{densityprediction} shows the stationary density distribution of dimers for three different values of $\omega$ obtained from simulations of Eq.~\eqref{Langevin}. The theoretical predictions of Eq.~\eqref{steady_state1} are in excellent agreement with the simulations. 

\begin{figure}[t]
\includegraphics[scale=0.33]{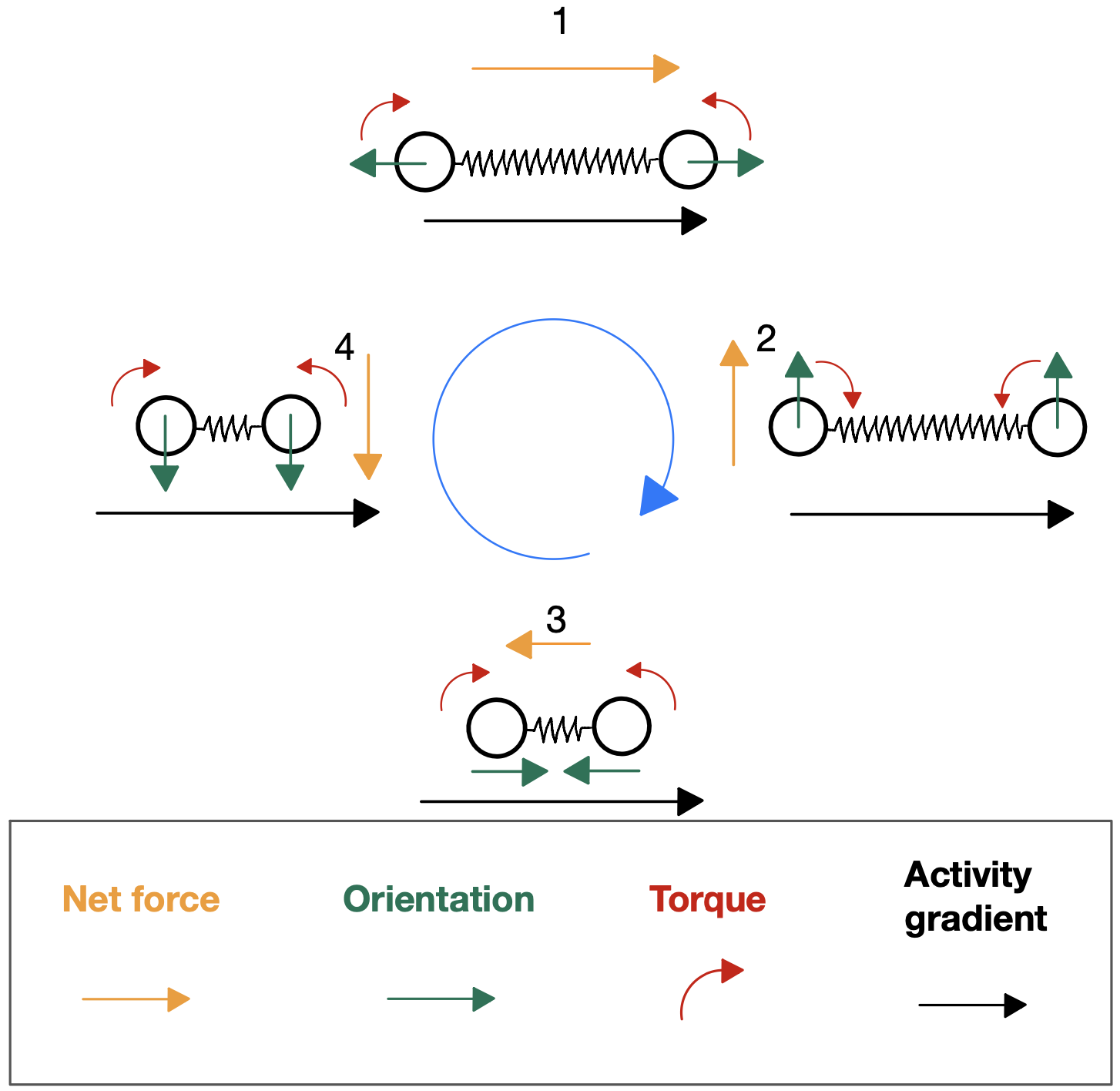}
\caption{Schematic representation of the mechanism leading to chemotaxis in the case of a dimer made by two active chiral particles with opposite and homogeneous torque (red arrows). The cycle above refers to the limiting case of large time scale related to rotational diffusion compared to the ones of the active torque and spring relaxation, i.e. $\tau \ll 1$ and $\Omega \gg 1$. The orientation vectors (green arrows) are typically opposite when the dimer is stretched (step $1$). For this reason, the dimer experiences a net force (orange arrow) pointing in the direction of the activity gradient (black arrow). The evolution of the orientation vectors due to the active torque leads, after some time $\sim 1/2\omega$, to the configuration in step $3$, with a net force towards the low activity region. However, since the spring in step $3$ is typically less extended than in $1$, this downwards force is much smaller than the one experienced by the dimer in step $1$. Overall, this asymmetry results in a biased motion toward high activity regions.
%Schematic representation of the dynamics of a dimer made by two active chiral particles with opposite and homogeneous torque (red arrows) starting from the initial condition in step $1$. This simplified mechanism is relative to the case $D_R\big( 1+D_R \tau\big) << \omega^2 \tau$ (negligible rotational diffusion) and assumes that the particle swaps ($3\rightarrow4$ and $6 \rightarrow 1$) take place on a shorter time scale than the one needed by the torque to rotate the orientation vector by $\frac{\pi}{2}$. The sum of the net forces (orange arrows) in a cycle points in the same direction of the activity gradient (black arrows).
}
\label{cycle}
\end{figure}

\begin{figure*}[t]
\includegraphics[scale=0.46
]{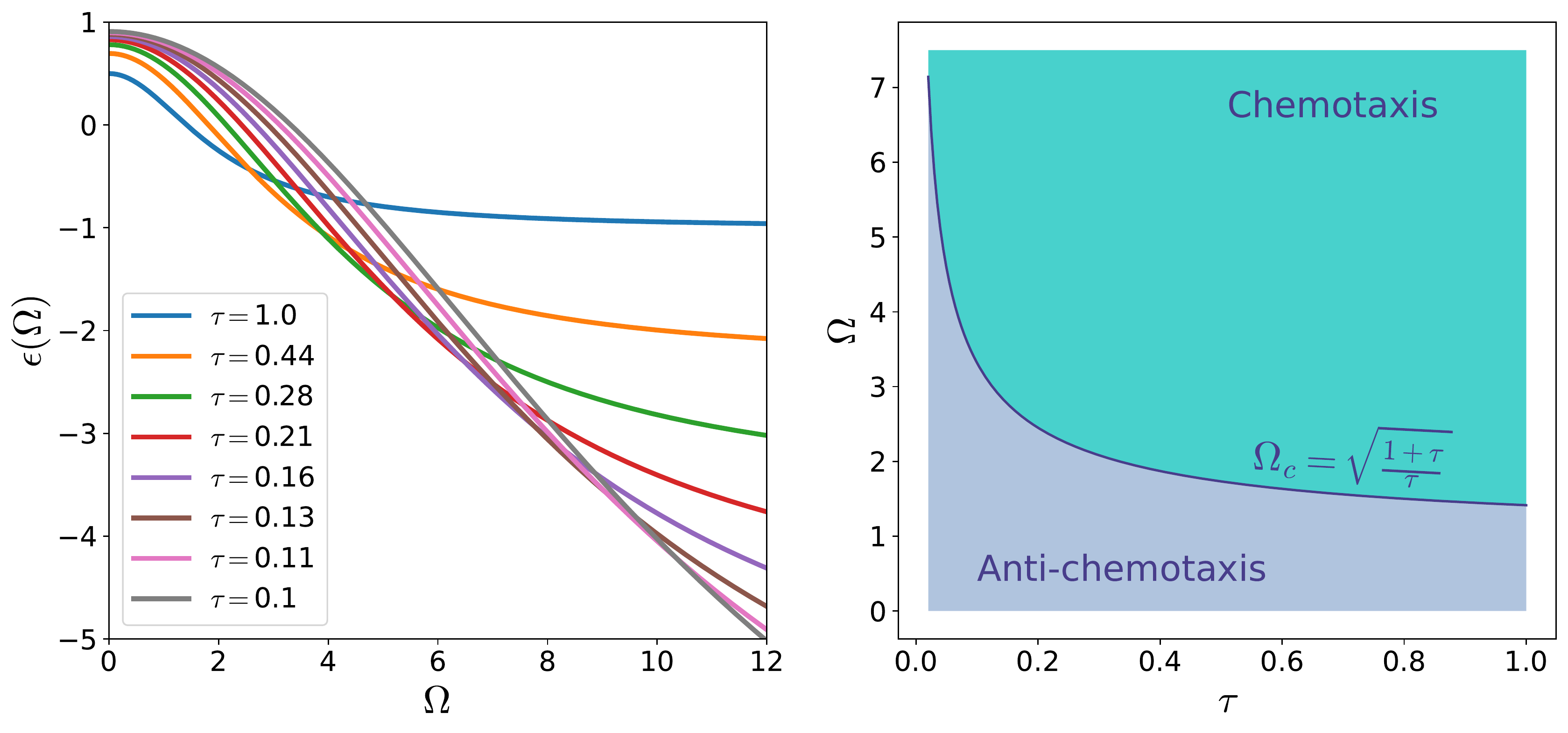}
\caption{Left: $\epsilon$ as a function of $\Omega$ for different values of $\tau$. For vanishing torque, the parameter $\epsilon \to \frac{1}{1+\tau}$, whereas for large torque $\epsilon \to -\frac{1}{\tau}$. With increasing active torque, a dimer switches behavior from antichemotactic ($\epsilon >0$) to chemotactic $\epsilon <0$. Right: Phase diagram in $(\Omega,\tau)$ plane. The blue critical curve separates the chemotactic from the anti-chemotactic region. In particular, for small spring relaxation time, it is much harder for the dimer to switch its tactic behaviour as the threshold torque tends to diverge. }
\label{phase_diagram}
\end{figure*}

At this stage the following question arises: what is the mechanism responsible for the emergence of a chemotactic phase and why does it require torque values above a certain threshold? To start addressing this point we focus on the limiting case where the time scale of rotational diffusion is much larger than the spring relaxation time scale, i.e., $\tau \ll 1$ and that of the active torque, i.e., $\Omega \gg 1$. We qualitatively describe the emergence of the chemotactic phase predicted by the model in this regime ($1+\tau << \Omega^2 \tau$) with the help of a schematic representation of the dynamics shown in Figure \ref{cycle}.
% effects of the rotational diffusion are negligible compared to those of the deterministic torque. We try to justify the emergence of the chemotactic phase predicted by the model in this regime ($D_R\big( 1+D_R \tau \big) << \omega^2 \tau$) with the help of a schematic representation of the dynamics reported in Figure \ref{cycle}.
Let us consider a configuration of the dimer in which one of the particles' orientation is in the direction of increasing activity whereas that of the other is in the direction of decreasing activity. In this configuration, the dimer is typically stretched and it experiences a net force towards the higher activity region. Accordingly, the dimer climbs up the activity gradient and the distance between the two particles increases. As the dimer climbs up, the orientations of the two particles evolve due to the active torque such that after some time ($\sim 1/2\omega$) they point towards each other. In this configuration, the dimer experiences a net force towards the low activity region. The dimer thus climbs down the activity gradient, however, while the dimer descends the activity gradient, the two particles approach each other such that the net downwards force continuously decreases before the two particles again point outwards as in the initial configuration. Overall, while climbing up the dimer experiences larger forces due to the two particles getting distant from each other than during the climb down when the two particles approach each other. In a cycle of period $1/\omega$, the dimer performs a forward and backward motion, with net drift towards the region of higher activities. We based our reasoning on a particular initial configuration. In general, if the rotational diffusion is negligible with respect to the deterministic torque, the mechanism will be somewhat different depending on the initial configuration, but in no case it will bias the motion toward region of lower activity. Note that for biased movement up the activity gradient only a sufficiently large active torque is necessary. The underlying mechanism does not require temporal integration of the fuel concentration~\cite{cates2012diffusive}, memory~\cite{kromer2020chemokinetic}, or an explicit coupling between the orientation of the particles and the activity gradients~\cite{lozano2016phototaxis,lozano2019diffusing,jahanshahi2020realization}.

The mechanism presented here relies on the changing distance between the two active particles. In the case of a rigid dimer where the distance between the two particles is fixed, there is no biased movement towards higher activity regions. We show analytically in Section $1$ of Supplementary Material that for a rigid dimer $\epsilon = 1/2$. Thus a rigid active chiral dimer always accumulates in the low activity regions. This is in contrast to the active-passive rigid dimers~\cite{vuijk2021chemotaxis}, where rigid dimers can also exhibit chemotaxis.

%However, since the two active particles approach each other in this configuration, 

% As can be seen from the figure, the sum of the net forces (orange arrows) in a cycle points toward the direction of the activity gradient. A necessary condition for this to happen is that the two (green) active forces maintain their relative orientation for each step of the cycle as sketched in the figure. For high values of $D_R$ the effects of stochasticity could compromise this condition and the effectiveness of the mechanism, as the time needed to decorrelate the relative orientation of the active forces decreases.
% %This is in agreement with the fact that the region below the critical curve in Figure \ref{phase_diagram} corresponds to the anti-chemotactic phase. 
% Notice that in the previous example we based our reasoning on a particular initial condition (step $1$). In general, if the rotational diffusion is negligible with respect to the deterministic torque, the mechanism will be somewhat different depending on the initial condition, but in no case it will bias the motion toward region of lower activity.
% Note that for biased movement up the activity gradient only a sufficiently large active torque is necessary. The underlying mechanism does not require temporal integration of the fuel concentration~\cite{cates2012diffusive}, memory~\cite{kromer2020chemokinetic}, or an explicit coupling between the orientation of the particles and the activity gradients~\cite{lozano2016phototaxis,lozano2019diffusing}.

Equation \eqref{critical_curve} defines the critical curve in the $(\Omega,\tau)$ plane separating the chemotactic phase from the anti-chemotactic one (See fig.~\ref{phase_diagram}). It is interesting to consider the limit $\tau \rightarrow 0$, in which case the critical torque scales as $\omega_c \sim \sqrt{kD_R}$. For a fixed $D_R$, the critical torque diverges in the limit of $k\rightarrow \infty$ implying that the dimer preferentially accumulates in the low activity regions. This is indeed expected as in this case, the two particles are tightly bound to each other and therefore, the dimer is effectively reduced to a single active particle which shows antichemotactic behavior. Similarly, for a fixed $k$, it is apparent that with increasing $D_R$, one requires increasingly large active torque to induce chemotactic behavior. 

It is important to note that the collective coordinate, $\bm{R}$, can describe the location of the dimer only when the two particles stay sufficiently close to each other. For small $k$, the activity gradients can become large on the length scale of the distance between the two particles. Our theory, based on the small-gradients approximation cannot describe such situation. 

\section{Odd diffusion}\label{ODD}

In the previous section we demonstrated how two interacting self propelled particles driven by microscopic opposite torques can exhibit, for sufficiently high torques, a chemotactic behavior.
The single components of this molecule belong to a class of systems defined as odd diffusive. This kind of systems, usually characterized by a break in time reversal and parity symmetry, has antisymmetric components in the diffusion tensor and flows perpendicular to the density gradient. Recently the analysis of odd diffusive systems has attracted a lot of attention for both equilibrium and out-of-equilibrium systems such as passive/active charged particles in magnetic field, see e.g. \cite{vuijk2020lorentz,abdoli2020nondiffusive,abdoli2020stationary,abdoli2020correlations}. In the case of charged self propelled (active) particles immersed in a magnetic field, rather than a microscopic driving torque, it is the Lorentz force that induces a certain chirality. Despite the different physical origin underlying their odd diffusive behavior, both self propelled particles (with torque induced chirality and Lorentz force induced chirality) are characterized by very similar dynamics. This similarity can be better appreciated from a comparison of their coarse grained FPEs (derived in Section $2$ and $3$ of Supplementary Material), where one integrates out the rotational degree of freedom. In particular, for a single ACP with spatially homogeneous driving torque $\omega$ and activity function $v_C$ in $d=2$, we have
\begin{equation}
\partial_t \rho(\bm{r},t)= \nabla_{\bm{r}} \cdot \Big[  \left( \frac{v^2_C}{2D_R} \underline{\underline{M}}^{-1} + D_T\bm{1}\right) \nabla_{\bm{r}}\rho(\bm{r},t) \Big]
\label{FPE_hACP}
\end{equation}
with $\underline{\underline{M}}=\bm{1}+\frac{\omega}{D_R}\bm{\varepsilon}$ and $\bm{\varepsilon}$ the totally antisymmetric Levi-Civita tensor in two dimensions. For a charged self-propelled particle immersed in homogeneous magnetic field $B$ and activity $v_B$ we have instead
\begin{equation}
    \partial_t \rho(\bm{r},t)=\nabla \cdot \left[ \left( \frac{v^2_B}{2D_R}  \frac{1}{1+\kappa^2} \bm{1} + D_T \underline{\underline{\Gamma}}^{-1} \right) \nabla \rho(\bm{r},t)\right],
    \label{FPE_hAPB}
\end{equation}
with $\gamma \underline{\underline{\Gamma}} = \gamma (\I - \kappa \bm{\varepsilon})$ the effective rank-$2$ friction tensor and $\kappa=\frac{qB}{\gamma}$. The two equations present a similar structure, with both diffusion tensors (expressions in round brackets) containing an anti-symmetric component. They can even be equivalent if the following relations hold 
\begin{equation}
    \begin{cases}
    v_C^2=-2 D_T \frac{D_R^2+\omega^2}{1+\kappa^2}\frac{\kappa}{\omega}\\
    v_B^2=2 D_R D_T \kappa \left( \kappa-\frac{D_R}{\omega} \right)
    \end{cases}.
\end{equation}
%In particular, the evolution of a single ACP with an active torque can be modeled in $d=2$ as
%\begin{equation}
%    \begin{split}
%    &\frac{d\bm{r}}{dt}=v_a(\bm{r})\bm{p}(\theta) + \sqrt{2 D_T} \bm{\xi}(t),\\
%    &\frac{d \theta}{dt}=\omega(\bm{r}) + \sqrt{2 D_R} \eta(t),
%    \end{split}
%\end{equation}
%with $v_a(\bm{r})$ the swim speed due to self propulsion along the direction $\bm{p}(\theta)$ and $\omega(\bm{r})$ the microscopic driving torque, whereas the underdamped evolution of a charged self-propelled particle under Lorentz force is given by
%\begin{equation}
%\begin{split}
%    &m\frac{d\bm{v}}{dt}=-\gamma\underline{\underline{\Gamma}}\bm{v}+f_a(\bm{r})\bm{p}(\theta) + \sqrt{2D_T}\bm{\xi}(t)\\
%    &\frac{d\bm{r}}{dt}=\bm{v}\\
%    &\frac{d \theta}{dt}= \sqrt{2 D_R} \eta(t),
%\end{split}
%\end{equation}
%where $f_a(\bm{r})\bm{p}(\theta)$ is the self propulsion force and the friction coefficient $\gamma$ becomes a rank-$2$ tensor $\gamma \underline{\underline{\Gamma}} = \gamma (\I - \kappa(\bm{r})\bm{\varepsilon})$  containing an anti-symmetric component, with $\bm{\varepsilon}$ the totally antisymmetric Levi-Civita tensor in two dimensions~\cite{vuijk2019anomalous,abdoli2020nondiffusive} and $\kappa(\bm{r})=\frac{q B(\bm{r})}{\gamma}$ . In the overdamped regime, it can be shown (see Supplementary Material, Section $2$ and $3$) that in the case of spatially homogeneous activity, magnetic field and active torque, the two systems can be mapped into each other.
Given this intriguing similarity between the two systems, it is natural to wonder whether this extends to their chemotactic behavior. In particular, can two self-propelled oppositely charged interacting particles immersed in a magnetic field $\bm{B}=B\hat{\bm{z}}$ cooperate in such a way as to exhibit chemotaxis, similarly to what has been demonstrated for a dimer composed of two active particles driven by opposite microscopic torques?\\
We show below that, despite the suggestive similarity to active chiral dimers, such charged dimers are always antichemotactic.
%We investigate whether this system is also characterized by the presence of two phases (chemotactic and anti-chemotactic) and, in that case, which parameters regulate their emergence.
To prove that, we need to work out an expression for the steady state density of the charged dimer. The derivation is similar to the one outlined in Section \ref{ACD} for a dimer composed of active particles driven by a torque (see Section $4$ of Supplementary Material for details) and starts from the evolution of the joint probability density:
\begin{equation}
    \partial_t P(\bm{r}_1,\bm{r}_2,\theta_1,\theta_2,t)=\left(\hat{\mathcal{L}}_{1}+\hat{\mathcal{L}}_{2}+\hat{\mathcal{L}}_{rot} \right)P
    \label{FPE_short_dimerB}
\end{equation}
with
\begin{equation}
\begin{split}
    &\hat{\mathcal{L}}_{1}=\nabla_{\bm{r}_1} \cdot \left[-\underline{\underline{\Gamma}}^{-1}_{+} \left(\frac{1}{\gamma}\bm{F} + v_a(\bm{r}_1) \bm{p}_1 +D_T  \nabla_{\bm{r}_1} \right) \right],\\
    \\
    &\hat{\mathcal{L}}_{2}=\nabla_{\bm{r}_2} \cdot \left[- \underline{\underline{\Gamma}}^{-1}_{-} \left(-\frac{1}{\gamma}\bm{F} + v_a(\bm{r}_2) \bm{p}_2 +D_T  \nabla_{\bm{r}_2} \right) \right],\\
    \\
    &\hat{\mathcal{L}}_{rot}=D_R (\frac{\partial^2}{\partial \theta_1^2} + \frac{\partial^2}{\partial \theta_2^2}).
    \end{split}
\end{equation}
The tensor $\underline{\underline{\Gamma}}_{\pm}$ has the same definition as $\underline{\underline{\Gamma}}$, with the additional subscript denoting the sign of the charge $\pm q$, $\bm{F}$ is the force experienced by the two monomers due to their attractive interaction and $v_a(\bm{r})$ is the activity function. Rewriting \eqref{FPE_short_dimerB} in terms of center-of-friction $\bm{R}=\frac{1}{2}(\underline{\underline{\Gamma}}_{+} \bm{r}_1 + \underline{\underline{\Gamma}}_{-} \bm{r}_2)$ and inner coordinate $\bm{r}=\bm{r}_1-\bm{r}_2$, and following similar steps to the ones presented for an active chiral molecule driven by opposite torques, we get a coarse grained FPE analogous to \eqref{VandD} with the following drift and diffusion terms 
\begin{equation}
    \begin{split}
    &D(\bm{R})=\frac{v_a^2(\bm{R})}{4D_R} + \frac{D_T}{2},\\
    \\
    &\bm{V}(\bm{R}) = -\frac{1}{2} \epsilon \nabla_{R} D(\bm{R}),\\
    \end{split}
\end{equation}
the exponent $\epsilon$ defined as:
\begin{equation}
    \epsilon=\left[1 - \frac{1-\kappa^2}{1+\kappa^2}\left(1-\frac{1}{1 + \tau(1 + \kappa^2)} \right) \right],
\end{equation}
and $\kappa = qB/\gamma$. By imposing again zero flux at steady state, the following density profile can be easily derived
\begin{equation}
    \rho(\bm{R}) \propto \left[1 + \frac{1}{D_T} \frac{v^2_a(\bm{R})}{2 D_R} \right]^{-\frac{\epsilon}{2}}.
\label{steady_state2}
\end{equation}
We note that the expressions \eqref{steady_state1} and \eqref{steady_state2} are equivalent when both the magnetic field $B$ and the torque $\omega$ are zero. Indeed, in this case both systems reduce to a dimer of simple ABPs. However, for generic non-zero values of $B$ and $\omega$, there is a striking difference between the two systems to be noticed. In the dimer of ACPs it is possible to change the sign of $\epsilon$ by varying the value of the torque $\omega$. This allows the system to explore both a chemotactic and an anti-chemotactic phase. Conversely, for the active dimer immersed in a magnetic field, the value of $\epsilon$ is strictly positive regardless the intensity of $B$. This constraint prevents the system from exhibiting a chemotactic phase.
\section{Discussion}\label{discussion_conclusions}
An active chiral particle preferentially accumulates in low activity regions. However, when two such particles are connected to each other with opposite chiralities, the resulting dimer can become chemotactic with increasing active torque. Such an assembly of active chiral particles might be possible using advanced fabrication techniques that have been used to assemble colloidal particles into desired structures~\cite{chen2011directed, bianchi2011patchy, sacanna2010lock, glotzer2007anisotropy, zhang2016directed, popescu2020chemically}. However, at the colloidal scale, it might be challenging to fabricate such dimers without affecting the rotation of the individual chiral particles. 

At the millimetre scale, our predictions could be tested in vibrots which are miniature robots that convert vibrations into rotational and translational motion ~\cite{altshuler2013vibrot, scholz2016ratcheting, scholz2017velocity, broseghini2019notched}. Vibrots with chiral leg configurations can rotate in both senses and hence mimic active chiral particles of opposite polarities~\cite{scholz2018rotating}. They can be connected to each other without affecting their rotation~\cite{scholz2021surfactants}. Activity gradients could be realised in a vibrating plate via surface patterning which modifies the effective friction experienced by a vibrot.

Active chiral particles belong to a class of systems referred to as odd-diffusive. In these systems, probability fluxes are not only along the density gradients but also perpendicular to them~\cite{hargus2021odd, kalz2022selfdiffusion, shinde2022strongly}. Odd-diffusive behavior is encoded in the diffusion tensor that has an antisymmetric part (Eq.~\eqref{FPE_hACP} and ~\eqref{FPE_hAPB}). We showed that despite sharing the property of odd-diffusion, the two systems, namely active chiral particles and Brownian particles under Lorentz force, exhibit distinct dynamics such that one cannot be mapped to another in general. We showed this explicitly in the context of the tactic behavior of dimers. Another example concerns the dynamics of charged Rouse dimers under magnetic field. These dimers exhibit strongly enhanced dynamics, even superballistic behavior, when subjected to a uniform magnetic field~\cite{shinde2022strongly}. The enhancement of dynamics predicted in charged Rouse dimers under magnetic field~\cite{shinde2022strongly} cannot be reproduced by dimers of active chiral particles. An example where the mapping is valid concerns the enhancement of self-diffusion due to interactions in odd-diffusive systems~\cite{kalz2022selfdiffusion}. This theoretical prediction can indeed be tested experimentally using active chiral particles. 

Since we considered particles with opposite chiralities, the dimer itself does not show odd-diffusive behavior. This is evident in the scalar valued diffusion coefficient in Eq.~\eqref{VandD}. Additional work is needed to characterize a dimer with arbitrary chiralities, which will, in general, exhibit odd-diffusive behavior. %It will also be interesting to study the dynamics of dimers in activity gradients.  
%However, the enhancement of dynamics predicted in charged Rouse dimers under magnetic field~\cite{shinde2022strongly} cannot be reproduced by dimers of active chiral particles.

% It was recently shown that collisions enhance the self-diffusion coefficient in odd-diffusive systems~\cite{kalz2022selfdiffusion}. 
% Charged Brownian particles under an external magnetic field constitutes the only equilibrium system that exhibits odd-diffusive behavior.

\bibliography{references}
% \appendix

% \section{Dimer of ACPs}
% \label{A}

% \section{Single active particle with Lorentz force}
% \label{B}

% \section{Single active chiral particle}
% \label{C}

% \section{Mapping}
% \label{D}

% \section{Active dimer under Lorentz force}
% \label{E}

\pagebreak
\widetext
\begin{center}
\textbf{\large Supplementary Material}
\end{center}
\setcounter{equation}{0}
\setcounter{section}{0}

\makeatletter
\renewcommand \thesection{S\@arabic\c@section}
\renewcommand\thetable{S\@arabic\c@table}
\renewcommand \thefigure{S\@arabic\c@figure}
\makeatother

\section*{Outline}
The present supplementary material contains the detailed derivation of the results reported in the main document. Section \ref{dimer_of_ACPs} shows how to derive the steady state density for a dimer made of two active chiral particles with opposite torque in a slowly varying activity gradient. The derivation is carried out with both rigid bond interaction and harmonic bond with zero rest length. Section \ref{sec_SACP} and \ref{sec_SAPB} deal with  single active chiral particles and single active particles under Lorentz force, respectively. In particular, for both systems, we show how to integrate out the orientation degree of freedom to get an effective FPE for the density distribution. Finally, section \ref{sec_dimerB} shows how to derive the steady state density for a dimer made of two oppositely charged active particles immersed in a magnetic field, again in the case of slowly varying activity function.

\section{Dimer of ACPs}
\label{dimer_of_ACPs}
The evolution of the system is described by the following dynamics:
\begin{equation}
    \begin{cases}
    \frac{d \bm{r}_1}{dt}=-\frac{1}{\gamma} \nabla_1 V(\bm{r}_1,\bm{r}_2) + v_a(\bm{r}_1) \bm{p}_1 + \sqrt{2 D_T} \bm{\xi}_1(t)\\
    \frac{d \bm{r}_2}{dt}=+ \frac{1}{\gamma} \nabla_1 V(\bm{r}_1,\bm{r}_2) + v_a(\bm{r}_2) \bm{p}_2 + \sqrt{2 D_T} \bm{\xi}_2(t)\\
    \frac{d \theta_1}{dt} = \omega + \sqrt{2 D_R} \eta_1(t) \\
    \frac{d \theta_2}{dt} = -\omega + \sqrt{2 D_R} \eta_2(t) \\
    \end{cases}
\end{equation}
where $\bm{p}_i=(\cos(\theta_i),\sin(\theta_i))$ and $\bm{\xi}_1(t),\bm{\xi}_2(t),\eta_1(t),\eta_2(t)$ are independent gaussian white noises
\begin{equation}
    \begin{cases}
    \braket{\xi_{i,j}(t)}=\braket{\eta_i(t)}=0\\
    \braket{\xi_{i,j}(t)\xi_{k,h}(s)}=\delta_{i,k}\delta_{j,h}\delta(t-s)\\
    \braket{\xi_{i,j}(t)\eta_k(s)}=0\\
    \braket{\eta_i(t)\eta_j(s)}=\delta_{i,j}\delta(t-s)\\
    \end{cases} i,j,k,h \in \{1,2\}
\end{equation}
For simplicity we will focus on the case of harmonic interaction potential $V(\bm{r}_1,\bm{r}_2)=\frac{k}{2}(\bm{r}_1-\bm{r}_2)^2$. We find convenient to perform a change of variables from $\{\bm{r}_1,\bm{r}_2 \}$ to $\{\bm{R}=\frac{\bm{r}_1+\bm{r}_2}{2}, \bm{r}=\bm{r}_1-\bm{r}_2 \}$, which we will refer to as center of mass and inner coordinates. The dynamics in terms of the new variables reads:
\begin{equation}
    \begin{cases}
    \frac{d \bm{R}}{dt}=\frac{1}{2} \left[ v_a(\bm{R}+\frac{\bm{r}}{2})\bm{p}_1+v_a(\bm{R}-\frac{\bm{r}}{2})\bm{p}_2  \right] + \sqrt{D_T} \bm{\xi}_1(t)\\
    \frac{d \bm{r}}{dt}= - 2 \frac{k}{\gamma}\bm{r} + \left[ v_a(\bm{R}+\frac{\bm{r}}{2})\bm{p}_1-v_a(\bm{R}-\frac{\bm{r}}{2})\bm{p}_2  \right] + \sqrt{4 D_T} \bm{\xi}_2(t)\\
    \frac{d \theta_1}{dt} = \omega + \sqrt{2 D_R} \eta_1(t) \\
    \frac{d \theta_2}{dt} = -\omega + \sqrt{2 D_R} \eta_2(t) \\
    \end{cases}
\end{equation}
To analyse the evolution of the joint probability density $P(\bm{R},\bm{r},\theta_1,\theta_2,t)$ we consider the Fokker Planck equation:
\begin{equation}
    \begin{split}
        \partial_t P(\bm{R},\bm{r},\theta_1,\theta_2,t) =& - \nabla_{\bm{R}} \cdot \Big[ \frac{1}{2}v_a \Big(\bm{R}+\frac{\bm{r}}{2} \Big)\bm{p}_1 P + \frac{1}{2}v_a \Big(\bm{R}-\frac{\bm{r}}{2}\Big)\bm{p}_2 P -\frac{D_T}{2} \nabla_{\bm{R}} P \Big] \\ &  -\nabla_{\bm{r}} \cdot \Big[ \frac{2}{\gamma} \bm{F}P + v_a \Big(\bm{R}+\frac{\bm{r}}{2} \Big)\bm{p}_1 P - v_a \Big(\bm{R}-\frac{\bm{r}}{2}\Big)\bm{p}_2 P - 2 D_T \nabla_{\bm{r}} P\Big]  \\& -\omega \frac{\partial P}{\partial \theta_1} +\omega \frac{\partial P}{\partial \theta_2}  + D_R \hat{\mathcal{R}}^2 P
    \end{split}
    \label{full_FP_1}
\end{equation}
where we defined the operator $\hat{\mathcal{R}}^2=\partial^2_{\theta_1}+\partial^2_{\theta_2}$. We would like to find an equation for the density as a function of the center of mass only. To do that we first expand $P(\bm{R},\bm{r},\theta_1,\theta_2,t)$ into the eigenfunctions of $\hat{\mathcal{R}}^2$ and we integrate out the orientational degrees of freedom (following the strategy presented in \cite{vuijk2021chemotaxis}).\\ We look for functions $\psi(\theta_1,\theta_2)$ solving $\hat{\mathcal{R}}^2 \psi(\theta_1,\theta_2)= \lambda \psi(\theta_1,\theta_2)$. Separating the variables $\psi(\theta_1,\theta_2)=\phi(\theta_1) \varphi(\theta_2)$ we get:
\begin{equation}
    \begin{cases}
    \partial^2_{\theta_1} \phi(\theta_1)=\lambda_1 \phi(\theta_1)\\
    \partial^2_{\theta_2} \varphi(\theta_2)=\lambda_2 \phi(\theta_2)\\
    \lambda=\lambda_1 + \lambda_2
    \end{cases}
\end{equation}
The two identical eigenproblems are satisfied by eigenfunctions of the form $\{1,\,\bm{p},\,\bm{p}\bm{p}-\frac{1}{2}\bm{I},...\}$ with eigenvalues $\{1,\,-1,\,-4,...\}$. Thus, $P(\bm{R},\bm{r},\theta_1,\theta_2,t)$ can be expanded as:
\begin{equation}
    P(\bm{R},\bm{r},\theta_1,\theta_2,t)=\frac{1}{\Omega_2^2} \left[\phi + \bm{\sigma}_1 \cdot \bm{p}_1 + \bm{\sigma}_2 \cdot \bm{p}_2 + \underline{\underline{\bm{\sigma}}}_{12} : \bm{p}_1\bm{p}_2 +  \underline{\underline{\bm{w}}}_1 :\left( \bm{p}_1\bm{p}_1 -\frac{1}{2}\bm{I} \right) +  \underline{\underline{\bm{w}}}_2 :\left( \bm{p}_2\bm{p}_2 -\frac{1}{2}\bm{I} \right) + \Theta\right]
\end{equation}
with $\Theta$ standing for higher order terms and $\Omega_2=2\pi$ being the $2$-dimensional solid angle.\\ The coefficients $(\phi, \bm{\sigma}_1,\bm{\sigma}_2,\underline{\underline{\bm{\sigma}}}_{12},\underline{\underline{\bm{w}}}_1,\underline{\underline{\bm{w}}}_2,...)$ in the previous expansion depend on both $\bm{R}$ and $\bm{r}$, and crucial information can be extracted from them. In particular, $\bm{\sigma}_i$ are related to the average orientation of the particles and $\underline{\underline{\bm{w}}}_i$ to the nematic order parameter \cite{de1993physics}. Moreover, we are aiming to find a FP equation
\begin{equation}
    \partial_t \rho(\bm{R})=-\nabla_{\bm{R}} \cdot \bm{J}=-\nabla_{\bm{R}} \cdot \left[ \bm{V}(\bm{R})\rho(\bm{R}) - D(\bm{R})\nabla_{\bm{R}}\rho(\bm{R})\right]
    \label{coarse_grained_FP}
\end{equation}
for the coarse grained density given by
\begin{equation}
    \rho(\bm{R})=\int d\bm{r} \phi(\bm{R},\bm{r})
\end{equation}
Since we are interested in the limit of small activity gradient, we will actually neglect the contributions coming from $\underline{\underline{\bm{w}}}_1$, $\underline{\underline{\bm{w}}}_2$ and $\Theta$ as they result in corrections to the flux $\bm{J}$ proportional to $\mathcal{O}(\nabla^2)$ (see \cite{vuijk2021chemotaxis} for a detailed justification of this approximation). Thus, we will work with the following truncated expansion for the joint probability density:
\begin{equation}
    P(\bm{R},\bm{r},\theta_1,\theta_2,t)=\frac{1}{\Omega_2^2} \left[\phi + \bm{\sigma}_1 \cdot \bm{p}_1 + \bm{\sigma}_2 \cdot \bm{p}_2 + \underline{\underline{\bm{\sigma}}}_{12} : \bm{p}_1\bm{p}_2 \right]
\end{equation}
After defining the scalar product
\begin{equation}
    \braket{f,g}=\int_{0}^{2\pi} d \theta_1 \int_{0}^{2\pi} d \theta_2 f(\theta_1,\theta_2)g(\theta_1,\theta_2)
\end{equation}
we can obtain a set of equations for the coefficients $\phi$,  $\bm{\sigma}_1$, $\bm{\sigma}_2$, $\underline{\underline{\bm{\sigma}}}_{12}$ by projecting the Fokker-Planck equations \eqref{full_FP_1} onto the eigenfunctions $1$, $\bm{p}_1$, $\bm{p}_2$ and $\bm{p}_1\bm{p}_2$. To this end, the following expressions will come helpful:
\begin{equation}
\begin{cases}
    \braket{1,P}=\phi\\
    \braket{p_{i,\alpha},P}=\frac{\sigma_{i,\alpha}}{2}\\
    \braket{p_{1,\alpha}p_{1,\beta},P}=\braket{p_{2,\alpha}p_{2,\beta},P}=\frac{\phi}{2}\delta_{\alpha,\beta}\\
    \braket{p_{1,\alpha}p_{2,\beta},P}=\frac{\sigma_{12,\alpha,\beta}}{4}\\
    \braket{1,\hat{\mathcal{R}}^2P}=0\\
    \braket{p_{i,\alpha},\hat{\mathcal{R}}^2P}=-\frac{\sigma_{i,\alpha}}{2}\\
    \braket{p_{1,\alpha}p_{2,\beta},\hat{\mathcal{R}}^2P}=-\frac{\sigma_{12,\alpha,\beta}}{2}\\
    \end{cases}
    \label{useful_avg_1}
\end{equation}
We will also need $\braket{p_{i,\alpha},\partial_{\theta_i}P}$, $\braket{p_{1,\alpha}p_{2,\beta},\partial_{\theta_1}P}$ and $\braket{p_{1,\alpha}p_{2,\beta},\partial_{\theta_2}P}$, whose expressions are explicitly worked out in the following.
\begin{equation}
\begin{split}
    \braket{p_{i,\alpha},\partial_{\theta_i}P}=\int d\theta_1 \, d\theta_2\, p_{i,\alpha}\partial_{\theta_i}P=-\int d\theta_1 \, d\theta_2\, \left( \partial_{\theta_i} p_{i,\alpha} \right)P
\end{split}
\end{equation}
Defining the matrix 
\begin{equation}
    \underline{\underline{U}}=\begin{pmatrix}
    0 & -1\\
    1 & 0
    \end{pmatrix}
\end{equation}
we can write:
\begin{equation}
\begin{split}
    \braket{p_{i,\alpha},\partial_{\theta_i}P}&=-\int d\theta_1 \, d\theta_2\,  U_{\alpha,\beta} p_{i,\beta} P= -U_{\alpha,\beta} \frac{\sigma_{i,\gamma}}{\Omega_2^2}\int d\theta_1 \, d\theta_2\, p_{i,\beta} p_{i,\gamma}=\\
    &=-U_{\alpha,\beta} \frac{\sigma_{i,\gamma}}{2} \delta_{\beta,\gamma}=-\frac{1}{2}U_{\alpha,\beta} \sigma_{i,\beta}
\end{split}
    \label{useful_avg_2}
\end{equation}
Analogously:
\begin{equation}
\begin{split}
    \braket{p_{1,\alpha}p_{2,\beta},\partial_{\theta_1}P}&=-\int d\theta_1 \, d\theta_2\,\left(\partial_{\theta_1}p_{1,\alpha}\right) p_{2,\beta}P=-\frac{1}{\Omega_2^2}U_{\alpha,\gamma}\int d\theta_1 \, d\theta_2\,p_{1,\gamma}\, p_{2,\beta}\, p_{1,h}\, p_{2,k}\, \sigma_{12,h,k}=\\
    &=-\frac{1}{4}U_{\alpha,\gamma}\delta_{\gamma,h}\delta_{\beta,k} \sigma_{12,h,k}=-\frac{1}{4}U_{\alpha,\gamma}\sigma_{12,\gamma,\beta}
    \end{split}
\label{useful_avg_3}
\end{equation}
and
\begin{equation}
\begin{split}
    \braket{p_{1,\alpha}p_{2,\beta},\partial_{\theta_2}P}&=-\int d\theta_1 \, d\theta_2\,p_{1,\alpha}\left(\partial_{\theta_2}p_{2,\beta}\right) P=-\frac{1}{\Omega_2^2}U_{\beta,\gamma}\int d\theta_1 \, d\theta_2\,p_{1,\alpha}\, p_{2,\gamma}\, p_{1,h}\, p_{2,k}\, \sigma_{12,h,k}=\\
    &=-\frac{1}{4}U_{\beta,\gamma}\delta_{\alpha,h}\delta_{\gamma,k} \sigma_{12,h,k}=-\frac{1}{4}U_{\beta,\gamma}\sigma_{12,\alpha,\gamma}
    \end{split}
\label{useful_avg_4}
\end{equation}
\paragraph{Equation for $\rho(\bm{R})$} By projecting \eqref{full_FP_1} onto the eigenfunction $1$ we get:
\begin{equation}
    \begin{split}
        \partial_t \phi(\bm{R},\bm{r},t) =& - \nabla_{\bm{R}} \cdot \Big[ \frac{1}{2}v_a \Big(\bm{R}+\frac{\bm{r}}{2} \Big)\frac{\bm{\sigma}_1}{2} + \frac{1}{2}v_a \Big(\bm{R}-\frac{\bm{r}}{2}\Big)\frac{\bm{\sigma}_2}{2} -\frac{D_T}{2} \nabla_{\bm{R}} \phi \Big] \\ &  -\nabla_{\bm{r}} \cdot \Big[ \frac{2}{\gamma} \bm{F}\phi + v_a \Big(\bm{R}+\frac{\bm{r}}{2} \Big)\frac{\bm{\sigma}_1}{2} - v_a \Big(\bm{R}-\frac{\bm{r}}{2}\Big)\frac{\bm{\sigma}_2}{2} - 2 D_T \nabla_{\bm{r}} \phi\Big] 
    \end{split}
\end{equation}
At this point, as we want to focus on the large scale behaviour of the dimer, we can integrate out the inner coordinate:
\begin{equation}
    \begin{split}
        \partial_t \rho(\bm{R},t) &= - \nabla_{\bm{R}} \cdot \Big[ \int d\bm{r}\, \frac{1}{4}v_a \Big(\bm{R}+\frac{\bm{r}}{2} \Big)\bm{\sigma}_1 + \int d\bm{r}\, \frac{1}{4}v_a \Big(\bm{R}-\frac{\bm{r}}{2}\Big)\bm{\sigma}_2 -\frac{D_T}{2} \nabla_{\bm{R}} \rho \Big]=\\
        &=- \nabla_{\bm{R}} \cdot \Big[\bm{J}_{\bm{\sigma}_1} + \bm{J}_{\bm{\sigma}_2} + \bm{J}_{D} \Big]
    \end{split}
    \label{rho_eq}
\end{equation}
In the last line we separated the contribution to the flux coming from the diffusion and the activity of the two monomers. To find an expression for the latter we need to analyze the evolution of the average polarizations $\bm{\sigma}_1$ and $\bm{\sigma}_2$.
\paragraph{Equation for $\bm{\sigma}_1$ and $\bm{\sigma}_2$}: Let us start considering the average polarization $\bm{\sigma}_1$. The equation describing its evolution can be found by projecting \eqref{full_FP_1} onto the eigenfunctions $p_{1,\alpha}$. By doing that (and denoting with $\partial_\alpha$ and $\partial'_\alpha$ respectively the $\alpha$-componenent of $\nabla_{\bm{R}}$ and $\nabla_{\bm{r}}$) we obtain:
\begin{equation}
    \begin{split}
        \frac{1}{2}\partial_t \sigma_{1,\alpha}(\bm{R},\bm{r},t) =& - \partial_{\beta} \Big[ \frac{1}{2}v_a \Big(\bm{R}+\frac{\bm{r}}{2} \Big)\braket{p_{1,\alpha}p_{1,\beta}, P} + \frac{1}{2}v_a \Big(\bm{R}-\frac{\bm{r}}{2}\Big)\braket{p_{1,\alpha}p_{2,\beta}, P} -\frac{D_T}{2} \partial_{\beta} \frac{\sigma_{1,\alpha}}{2} \Big] \\ &  -\partial'_{\beta} \Big[ \frac{2}{\gamma} F_{\beta}\frac{\sigma_{1,\alpha}}{2} + v_a \Big(\bm{R}+\frac{\bm{r}}{2} \Big)\braket{p_{1,\alpha}p_{1,\beta}, P} - v_a \Big(\bm{R}-\frac{\bm{r}}{2}\Big)\braket{p_{1,\alpha}p_{2,\beta}, P} - 2 D_T \partial'_{\beta} \frac{\sigma_{1,\alpha}}{2}\Big]  \\& -\omega \braket{p_{1,\alpha}, \partial _{\theta_1} P} +\omega \braket{p_{1,\alpha}, \partial _{\theta_2} P}  + D_R \braket{p_{1,\alpha},\hat{\mathcal{R}}^2 P}
    \end{split}
\end{equation}
With the help of \eqref{useful_avg_1} and \eqref{useful_avg_2} we have:
\begin{equation}
    \begin{split}
        \partial_t \sigma_{1,\alpha}(\bm{R},\bm{r},t) =& - \partial_{\beta} \Big[ \frac{1}{2}v_a \Big(\bm{R}+\frac{\bm{r}}{2} \Big)\delta_{\alpha,\beta}\phi + \frac{1}{2}v_a \Big(\bm{R}-\frac{\bm{r}}{2}\Big)\frac{\sigma_{12,\alpha,\beta}}{2} -\frac{D_T}{2} \partial_{\beta} \sigma_{1,\alpha} \Big] \\ &  -\partial'_{\beta} \Big[ \frac{2}{\gamma} F_{\beta}\sigma_{1,\alpha} + v_a \Big(\bm{R}+\frac{\bm{r}}{2} \Big)\delta_{\alpha,\beta}\phi - v_a \Big(\bm{R}-\frac{\bm{r}}{2}\Big)\frac{\sigma_{12,\alpha,\beta}}{2} - 2 D_T \partial'_{\beta} \sigma_{1,\alpha}\Big]  \\& +\omega U_{\alpha,\beta} \sigma_{1,\beta} - D_R \sigma_{1,\alpha}
    \end{split}
\end{equation}
In vector notation:
\begin{equation}
    \begin{split}
        \partial_t \bm{\sigma}_{1}(\bm{R},\bm{r},t) =& - \nabla_{\bm{R}} \cdot \Big[ \frac{1}{2}v_a \Big(\bm{R}+\frac{\bm{r}}{2} \Big)\phi \bm{1} + \frac{1}{4}v_a \Big(\bm{R}-\frac{\bm{r}}{2}\Big)\underline{\underline{\bm{\sigma}}}^T_{12} -\frac{D_T}{2} \nabla_{\bm{R}} \bm{\sigma}_{1} \Big] \\ &  -\nabla_{\bm{r}} \cdot \Big[ \frac{2}{\gamma} \bm{F}\bm{\sigma}_{1} + v_a \Big(\bm{R}+\frac{\bm{r}}{2} \Big)\phi \bm{1} - \frac{1}{2}v_a \Big(\bm{R}-\frac{\bm{r}}{2}\Big)\underline{\underline{\bm{\sigma}}}^T_{12} - 2 D_T \nabla_{\bm{r}} \bm{\sigma}_{1}\Big]  \\& - \left( D_R\bm{1}-\omega \underline{\underline{U}}    \right)\bm{\sigma}_1
    \end{split}
\end{equation}
A similar derivation for the average polarization $\bm{\sigma}_2$ leads to:
\begin{equation}
    \begin{split}
        \partial_t \bm{\sigma}_{2}(\bm{R},\bm{r},t) =& - \nabla_{\bm{R}} \cdot \Big[ \frac{1}{2}v_a \Big(\bm{R}-\frac{\bm{r}}{2} \Big)\phi \bm{1} + \frac{1}{4}v_a \Big(\bm{R}+\frac{\bm{r}}{2}\Big)\underline{\underline{\bm{\sigma}}}_{12} -\frac{D_T}{2} \nabla_{\bm{R}} \bm{\sigma}_{2} \Big] \\ &  -\nabla_{\bm{r}} \cdot \Big[ \frac{2}{\gamma} \bm{F}\bm{\sigma}_{2} - v_a \Big(\bm{R}-\frac{\bm{r}}{2} \Big)\phi \bm{1} + \frac{1}{2}v_a \Big(\bm{R}+\frac{\bm{r}}{2}\Big)\underline{\underline{\bm{\sigma}}}_{12} - 2 D_T \nabla_{\bm{r}} \bm{\sigma}_{2}\Big]  \\& - \left( D_R\bm{1}+\omega \underline{\underline{U}}    \right)\bm{\sigma}_2
    \end{split}
\end{equation}
\paragraph{Equation for $\underline{\underline{\sigma}}_{12}$:} Finally, an equation for the parameter $\underline{\underline{\sigma}}_{12}$ can be derived by projecting \eqref{full_FP_1} onto the eigenfunctions $p_{1,\alpha}p_{2,\beta}$. This is readily done in the following:
\begin{equation}
    \begin{split}
        \frac{1}{4}\partial_t \sigma_{12,\alpha,\beta} =& - \partial_{\gamma} \Big[ \frac{1}{2}v_a \Big(\bm{R}+\frac{\bm{r}}{2} \Big)\braket{p_{1,\alpha}p_{2,\beta}p_{1,\gamma}, P} + \frac{1}{2}v_a \Big(\bm{R}-\frac{\bm{r}}{2}\Big)\braket{p_{1,\alpha}p_{2,\beta}p_{2,\gamma}, P} -\frac{D_T}{2} \partial_{\gamma} \frac{\sigma_{12,\alpha,\beta}}{4} \Big] \\ &  -\partial'_{\gamma} \Big[ \frac{2}{\gamma} F_{\gamma}\frac{\sigma_{12,\alpha,\beta}}{4} + v_a \Big(\bm{R}+\frac{\bm{r}}{2} \Big)\braket{p_{1,\alpha}p_{2,\beta}p_{1,\gamma}, P} - v_a \Big(\bm{R}-\frac{\bm{r}}{2}\Big)\braket{p_{1,\alpha}p_{2,\beta}p_{2,\gamma}, P} - 2 D_T \partial'_{\gamma} \frac{\sigma_{12,\alpha,\beta}}{4}\Big]  \\& -\omega \braket{p_{1,\alpha}p_{2,\beta}, \partial _{\theta_1} P} +\omega \braket{p_{1,\alpha}p_{2,\beta}, \partial _{\theta_2} P}  + D_R \braket{p_{1,\alpha}p_{2,\beta},\hat{\mathcal{R}}^2 P}
    \end{split}
    \label{eq_sigma12}
\end{equation}
Before to go on we actually need to compute the scalar products $\braket{p_{1,\alpha}p_{2,\beta}p_{1,\gamma}, P}$ and $\braket{p_{1,\alpha}p_{2,\beta}p_{2,\gamma}, P}$. The first one reads:
\begin{equation}
    \braket{p_{1,\alpha}p_{2,\beta}p_{1,\gamma}, P}=\frac{1}{\Omega_2^2}\sigma_{2,\delta}\braket{p_{1,\alpha}p_{2,\beta}p_{1,\gamma},p_{2,\delta}}=\frac{1}{4}\sigma_{2,\delta}\delta_{\alpha,\gamma} \delta_{\beta,\delta}=\frac{1}{4}\sigma_{2,\beta}\delta_{\alpha,\gamma} 
\end{equation}
Similarly, the second one:
\begin{equation}
    \braket{p_{1,\alpha}p_{2,\beta}p_{2,\gamma}, P}=\frac{1}{\Omega_2^2}\sigma_{1,\delta}\braket{p_{1,\alpha}p_{2,\beta}p_{2,\gamma},p_{1,\delta}}=\frac{1}{4}\sigma_{1,\delta}\delta_{\alpha,\delta} \delta_{\beta,\gamma}=\frac{1}{4}\sigma_{1,\alpha}\delta_{\beta,\gamma} 
\end{equation}
Thus, plugging these expressions back in \eqref{eq_sigma12} together with \eqref{useful_avg_1}, \eqref{useful_avg_3} and \eqref{useful_avg_4}, we get: 
\begin{equation}
    \begin{split}
        \partial_t \sigma_{12,\alpha,\beta} =& - \partial_{\gamma} \Big[ \frac{1}{2}v_a \Big(\bm{R}+\frac{\bm{r}}{2} \Big)\sigma_{2,\beta}\delta_{\alpha,\gamma}  + \frac{1}{2}v_a \Big(\bm{R}-\frac{\bm{r}}{2}\Big)\sigma_{1,\alpha}\delta_{\beta,\gamma}  -\frac{D_T}{2} \partial_{\gamma} \sigma_{12,\alpha,\beta} \Big] \\ &  -\partial'_{\gamma} \Big[ \frac{2}{\gamma} F_{\gamma}\sigma_{12,\alpha,\beta} + v_a \Big(\bm{R}+\frac{\bm{r}}{2} \Big)\sigma_{2,\beta}\delta_{\alpha,\gamma}  - v_a \Big(\bm{R}-\frac{\bm{r}}{2}\Big)\sigma_{1,\alpha}\delta_{\beta,\gamma}  - 2 D_T \partial'_{\gamma} \sigma_{12,\alpha,\beta}\Big]  \\& +\omega U_{\alpha,\gamma}\sigma_{12,\gamma,\beta} -\omega U_{\beta,\gamma}\sigma_{12,\alpha,\gamma}  -2 D_R \sigma_{12,\alpha,\beta}
    \end{split}
\end{equation}
In vector notation:
\begin{equation}
    \begin{split}
        \partial_t \underline{\underline{\bm{\sigma}}}_{12} =& -\frac{1}{2} \underline{\underline{J}}^T_{\bm{R}} \left[ v_a \Big(\bm{R}+\frac{\bm{r}}{2} \Big)\bm{\sigma}_{2}\right] - \frac{1}{2} \underline{\underline{J}}_{\bm{R}} \left[v_a \Big(\bm{R}-\frac{\bm{r}}{2}\Big)\bm{\sigma}_{1}\right]  +\frac{D_T}{2} \nabla^2_{\bm{R}} \underline{\underline{\bm{\sigma}}}_{12} + \\ &  -\frac{2}{\gamma} \nabla_{\bm{r}} \cdot \left[  \bm{F}\underline{\underline{\bm{\sigma}}}_{12} \right] - \underline{\underline{J}}^T_{\bm{r}} \left[ v_a \Big(\bm{R}+\frac{\bm{r}}{2} \Big)\bm{\sigma}_{2}\right]  + \underline{\underline{J}}_{\bm{R}} \left[v_a \Big(\bm{R}-\frac{\bm{r}}{2}\Big)\bm{\sigma}_{1} \right] + 2 D_T \nabla^2_{\bm{r}} \underline{\underline{\bm{\sigma}}}_{12}+  \\& +\omega \underline{\underline{U}}\,\underline{\underline{\bm{\sigma}}}_{12} +\omega \underline{\underline{\bm{\sigma}}}_{12}\,\underline{\underline{U}}  -2 D_R \underline{\underline{\bm{\sigma}}}_{12}
    \end{split}
\end{equation}
where we denoted with $\underline{\underline{J}}_{\bm{R}}[\cdot]$ and $\underline{\underline{J}}_{\bm{r}}[\cdot]$ the Jacobian matrices relative to the center of mass and the inner variable. Notice that the last line can be rewritten as $-\bm{T}:\underline{\underline{\bm{\sigma}}}_{12}$, where $\bm{T}$ is a suitable invertible rank-$4$ tensor and $:$ indicates the double contraction. \\
The equation \eqref{rho_eq} for the density $\rho$ has the form of a continuity equation, meaning that it is a locally conserved quantity and its relaxation depends on the spatial gradients characterizing the system. Conversely, the sink terms appearing in the equations for the average polarizations $\bm{\sigma}_1$, $\bm{\sigma}_2$ and for the matrix $\underline{\underline{\bm{\sigma}}}_{12}$ reveal that they relax on a much faster time scale. For this reason we will treat $\bm{\sigma}_1$, $\bm{\sigma}_2$ and $\underline{\underline{\bm{\sigma}}}_{12}$ as fast modes, and we will neglect the temporal derivative in their equations. Thus, we can rewrite our hierarchical set of equations as:
\begin{equation}
\begin{cases}
\partial_t \rho(\bm{R},t) = - \nabla_{\bm{R}} \cdot \Big[ \int d\bm{r}\, \frac{1}{4}v_a \Big(\bm{R}+\frac{\bm{r}}{2} \Big)\bm{\sigma}_1 + \int d\bm{r}\, \frac{1}{4}v_a \Big(\bm{R}-\frac{\bm{r}}{2}\Big)\bm{\sigma}_2 -\frac{D_T}{2} \nabla_{\bm{R}} \rho \Big]\\  
\\
\begin{split}
        \bm{\sigma}_1=& - \left( D_R\bm{1}-\omega \underline{\underline{U}}    \right)^{-1} \Big\{ \nabla_{\bm{R}} \cdot \Big[ \frac{1}{2}v_a \Big(\bm{R}+\frac{\bm{r}}{2} \Big)\phi \bm{1} + \frac{1}{4}v_a \Big(\bm{R}-\frac{\bm{r}}{2}\Big)\underline{\underline{\bm{\sigma}}}^T_{12} -\frac{D_T}{2} \nabla_{\bm{R}} \bm{\sigma}_{1} \Big] +\\ &  +\nabla_{\bm{r}} \cdot \Big[ \frac{2}{\gamma} \bm{F}\bm{\sigma}_{1} + v_a \Big(\bm{R}+\frac{\bm{r}}{2} \Big)\phi \bm{1} - \frac{1}{2}v_a \Big(\bm{R}-\frac{\bm{r}}{2}\Big)\underline{\underline{\bm{\sigma}}}^T_{12} - 2 D_T \nabla_{\bm{r}} \bm{\sigma}_{1}\Big] \Big\}
    \end{split}
\\
\\
    \begin{split}
        \bm{\sigma}_2=& - \left( D_R\bm{1}+\omega \underline{\underline{U}}    \right)^{-1} \Big\{ \nabla_{\bm{R}} \cdot \Big[ \frac{1}{2}v_a \Big(\bm{R}-\frac{\bm{r}}{2} \Big)\phi \bm{1} + \frac{1}{4}v_a \Big(\bm{R}+\frac{\bm{r}}{2}\Big)\underline{\underline{\bm{\sigma}}}_{12} -\frac{D_T}{2} \nabla_{\bm{R}} \bm{\sigma}_{2} \Big] \\ &  +\nabla_{\bm{r}} \cdot \Big[ \frac{2}{\gamma} \bm{F}\bm{\sigma}_{2} - v_a \Big(\bm{R}-\frac{\bm{r}}{2} \Big)\phi \bm{1} + \frac{1}{2}v_a \Big(\bm{R}+\frac{\bm{r}}{2}\Big)\underline{\underline{\bm{\sigma}}}_{12} - 2 D_T \nabla_{\bm{r}} \bm{\sigma}_{2}\Big]\Big\}
    \end{split}
\\
\\
 \begin{split}
        \underline{\underline{\bm{\sigma}}}_{12} =& \bm{T}^{-1}:\Big\{ -\frac{1}{2} \underline{\underline{J}}^T_{\bm{R}} \left[ v_a \Big(\bm{R}+\frac{\bm{r}}{2} \Big)\bm{\sigma}_{2}\right] - \frac{1}{2} \underline{\underline{J}}_{\bm{R}} \left[v_a \Big(\bm{R}-\frac{\bm{r}}{2}\Big)\bm{\sigma}_{1}\right]  +\frac{D_T}{2} \nabla^2_{\bm{R}} \underline{\underline{\bm{\sigma}}}_{12} + \\ &  -\frac{2}{\gamma} \nabla_{\bm{r}} \cdot \left[  \bm{F}\underline{\underline{\bm{\sigma}}}_{12} \right] - \underline{\underline{J}}^T_{\bm{r}} \left[ v_a \Big(\bm{R}+\frac{\bm{r}}{2} \Big)\bm{\sigma}_{2}\right]  + \underline{\underline{J}}_{\bm{R}} \left[v_a \Big(\bm{R}-\frac{\bm{r}}{2}\Big)\bm{\sigma}_{1} \right] + 2 D_T \nabla^2_{\bm{r}} \underline{\underline{\bm{\sigma}}}_{12} \Big\}
    \end{split}
\end{cases}
\end{equation}
Before to proceed with the analysis of the fluxes $\bm{J}_{
\bm{\sigma}_1}$ and $\bm{J}_{
\bm{\sigma}_2}$, a further simplification needs to be done at this stage. As anticipated, being interested in the case of slowly varying swim force, we consider $v_a (\bm{R})$ with gradients much smaller than the persistence length induced by the activity. For this reason, we will neglect all the contributions to the flux of $\rho(\bm{R})$ which are proportional to $\mathcal{O}(\nabla^2)$ \cite{vuijk2021chemotaxis}. In particular, thanks to this approximation, the matrix $\underline{\underline{\bm{\sigma}}}_{12}$ will play no role in the rest of the derivation. \\
At this point, we are ready to focus on the density flux, starting with $\bm{J}_{
\bm{\sigma}_1}$:
\begin{equation}
\begin{split}
    \bm{J}_{
\bm{\sigma}_1}&=\int d\bm{r}\, \frac{1}{4}v_a \Big(\bm{R}+\frac{\bm{r}}{2} \Big)\bm{\sigma}_1=\\
&=-\int d\bm{r}\, \frac{1}{4}v_a \Big(\bm{R}+\frac{\bm{r}}{2} \Big) \left( D_R\bm{1}-\omega \underline{\underline{U}}    \right)^{-1} \Big\{ \nabla_{\bm{R}} \cdot \Big[ \frac{1}{2}v_a \Big(\bm{R}+\frac{\bm{r}}{2} \Big)\phi \bm{1}\Big] +\nabla_{\bm{r}} \cdot \Big[ \frac{2}{\gamma} \bm{F}\bm{\sigma}_{1} + v_a \Big(\bm{R}+\frac{\bm{r}}{2} \Big)\phi \bm{1}\Big] \Big\}=\\
&=\left( D_R\bm{1}-\omega \underline{\underline{U}}    \right)^{-1} \Big\{-\int d\bm{r}\, \frac{1}{8}v_a \Big(\bm{R}+\frac{\bm{r}}{2} \Big) \nabla_{\bm{R}} \cdot \Big[v_a \Big(\bm{R}+\frac{\bm{r}}{2} \Big)\phi \bm{1}\Big] + \int d\bm{r}\, \frac{1}{8} \left[ \nabla_{\bm{R}}  v_a \Big(\bm{R}+\frac{\bm{r}}{2} \Big)\right] v_a \Big(\bm{R}+\frac{\bm{r}}{2} \Big)\phi + \\
&+\int d\bm{r}\, \frac{1}{4 \gamma} \left[ \nabla_{\bm{R}}  v_a \Big(\bm{R}+\frac{\bm{r}}{2} \Big)\right] \cdot \bm{F}\bm{\sigma}_{1} \Big\}
\end{split}
\label{flux_sigma1}
\end{equation}
where we integrated by parts the second and third integral and used $\nabla_{\bm{r}} v_a \Big(\bm{R}+\frac{\bm{r}}{2} \Big)=\frac{1}{2}\nabla_{\bm{R}} v_a \Big(\bm{R}+\frac{\bm{r}}{2} \Big)$. Let us now consider the last integral separately:
\begin{equation}
\begin{split}
    \bm{I}_1&=\int d\bm{r}\, \frac{1}{4 \gamma} \left[ \nabla_{\bm{R}}  v_a \Big(\bm{R}+\frac{\bm{r}}{2} \Big)\cdot \bm{F}\right] \bm{\sigma}_{1}=\\
    &=-\frac{1}{4 \gamma}\left( D_R\bm{1}-\omega \underline{\underline{U}} \right)^{-1} \int d\bm{r}\,\left[ \nabla_{\bm{R}}  v_a \Big(\bm{R}+\frac{\bm{r}}{2} \Big)\cdot \bm{F}\right] \nabla_{\bm{r}} \cdot \Big[ \frac{2}{\gamma} \bm{F}\bm{\sigma}_{1} + v_a \Big(\bm{R}+\frac{\bm{r}}{2} \Big)\phi \bm{1}\Big]=\\
    &=+\frac{1}{4 \gamma}\left( D_R\bm{1}-\omega \underline{\underline{U}} \right)^{-1} \int d\bm{r}\, \nabla_{\bm{r}} \left[ \nabla_{\bm{R}}  v_a \Big(\bm{R}+\frac{\bm{r}}{2} \Big)\cdot \bm{F}\right] \cdot \Big[ \frac{2}{\gamma} \bm{F}\bm{\sigma}_{1} + v_a \Big(\bm{R}+\frac{\bm{r}}{2} \Big)\phi \bm{1}\Big]=\\
    &=+\frac{1}{4 \gamma}\left( D_R\bm{1}-\omega \underline{\underline{U}} \right)^{-1} \int d\bm{r}\,  \left[ \left(\nabla_{\bm{r}} \bm{F} \right)\cdot \nabla_{\bm{R}}  v_a \Big(\bm{R}+\frac{\bm{r}}{2} \Big) \right] \cdot \Big[ \frac{2}{\gamma} \bm{F}\bm{\sigma}_{1} + v_a \Big(\bm{R}+\frac{\bm{r}}{2} \Big)\phi \bm{1}\Big]
\end{split}
\label{I_1}
\end{equation}
where in the last line we neglected the term proportional to $\nabla_{\bm{r}}\nabla_{\bm{R}}v_a \Big(\bm{R}+\frac{\bm{r}}{2} \Big)=\mathcal{O}(\nabla^2_{\bm{R}})$. For a spring with zero rest length and stiffness $k$, we have $\nabla_{\bm{r}}\bm{F}=-k\bm{1}$ (the derivation could be easily generalized to the case of a spring with finite rest length). Thus:
\begin{equation}
\begin{split}
    \bm{I}_1&=-\frac{k}{4 \gamma}\left( D_R\bm{1}-\omega \underline{\underline{U}} \right)^{-1} \int d\bm{r}\,  \left[ \nabla_{\bm{R}}  v_a \Big(\bm{R}+\frac{\bm{r}}{2} \Big) \right] \cdot \Big[ \frac{2}{\gamma} \bm{F}\bm{\sigma}_{1} + v_a \Big(\bm{R}+\frac{\bm{r}}{2} \Big)\phi \bm{1}\Big]=\\
    &=-\frac{2k}{\gamma} \left( D_R\bm{1}-\omega \underline{\underline{U}} \right)^{-1} \Big\{ \bm{I}_1 + \frac{1}{8} \int d\bm{r}\,v_a \Big(\bm{R}+\frac{\bm{r}}{2} \Big) \phi \nabla_{\bm{R}}  v_a \Big(\bm{R}+\frac{\bm{r}}{2} \Big) \Big\}
\end{split}
\end{equation}
from which
\begin{equation}
    \bm{I}_1=-\frac{1}{8}\left[ \left( 1+\frac{\gamma D_R}{2 k}\right)\bm{1}-\frac{\gamma \omega}{2 k} \underline{\underline{U}} \right]^{-1} \int d\bm{r}\,v_a \Big(\bm{R}+\frac{\bm{r}}{2} \Big) \phi \nabla_{\bm{R}}  v_a \Big(\bm{R}+\frac{\bm{r}}{2} \Big)
\end{equation}
Plugging this equation back in \eqref{flux_sigma1} we have:
\begin{equation}
\begin{split}
        \bm{J}_{
\bm{\sigma}_1}&=\left( D_R\bm{1}-\omega \underline{\underline{U}}    \right)^{-1} \Big\{-\int d\bm{r}\, \frac{1}{8}v_a \Big(\bm{R}+\frac{\bm{r}}{2} \Big) \nabla_{\bm{R}} \cdot \Big[ v_a \Big(\bm{R}+\frac{\bm{r}}{2} \Big)\phi \bm{1}\Big] + \int d\bm{r}\, \frac{1}{8} \left[ \nabla_{\bm{R}}  v_a \Big(\bm{R}+\frac{\bm{r}}{2} \Big)\right] v_a \Big(\bm{R}+\frac{\bm{r}}{2} \Big)\phi + \\
& -\frac{1}{8}\left[ \left( 1+\frac{\gamma D_R}{2 k}\right)\bm{1}-\frac{\gamma \omega}{2 k} \underline{\underline{U}} \right]^{-1} \int d\bm{r}\,v_a \Big(\bm{R}+\frac{\bm{r}}{2} \Big) \phi \nabla_{\bm{R}}  v_a \Big(\bm{R}+\frac{\bm{r}}{2} \Big)\Big\}
\end{split}
\end{equation}
Since the distance between the two monomers is assumed to be small compared to the activity gradients, we approximate $\phi(\bm{\bm{R}},\bm{\bm{r}},t) \approx \rho(\bm{\bm{R}},t)\delta(\bm{r})$ inside the integrals, getting:
\begin{equation}
\begin{split}
        \bm{J}_{
\bm{\sigma}_1}&=\left( D_R\bm{1}-\omega \underline{\underline{U}}    \right)^{-1} \Big\{- \frac{1}{8}v^2_a \left(\bm{R}\right) \nabla_{\bm{R}}  \rho(\bm{R}) -\frac{1}{16} \rho(\bm{R})\left[ \left( 1+\frac{\gamma D_R}{2 k}\right)\bm{1}-\frac{\gamma \omega}{2 k} \underline{\underline{U}} \right]^{-1}  \nabla_{\bm{R}}  v^2_a \left(\bm{R}\right)\Big\}
\end{split}
\end{equation}
An analogous reasoning for $\bm{J}_{
\bm{\sigma}_2}$ leads to:
\begin{equation}
\begin{split}
        \bm{J}_{
\bm{\sigma}_2}&=\left( D_R\bm{1}+\omega \underline{\underline{U}}    \right)^{-1} \Big\{- \frac{1}{8}v^2_a \left(\bm{R}\right) \nabla_{\bm{R}}  \rho(\bm{R}) -\frac{1}{16} \rho(\bm{R})\left[ \left( 1+\frac{\gamma D_R}{2 k}\right)\bm{1}+\frac{\gamma \omega}{2 k} \underline{\underline{U}} \right]^{-1}  \nabla_{\bm{R}}  v^2_a \left(\bm{R}\right)\Big\}
\end{split}
\end{equation}
To sum the two fluxes we first need to compute the following quantities:
\begin{equation}
    \left( D_R\bm{1}-\omega \underline{\underline{U}}\right)^{-1} + \left( D_R\bm{1}+\omega \underline{\underline{U}} \right)^{-1}=\frac{1}{D^2_R+\omega^2}\left[ \begin{pmatrix}D_R & -\omega\\
    \omega & D_R
    \end{pmatrix} + \begin{pmatrix}D_R & +\omega\\
    -\omega & D_R
    \end{pmatrix}
    \right]=\frac{2D_R}{D^2_R+\omega^2} \bm{1}
    \label{matrix_1}
\end{equation}
\begin{equation}
\begin{split}
    &\left( D_R\bm{1}-\omega \underline{\underline{U}}\right)^{-1} \left[ \left( 1+\frac{\gamma D_R}{2 k}\right)\bm{1}-\frac{\gamma \omega}{2 k} \underline{\underline{U}} \right]^{-1}=\left\{ \left[ \left( 1+\frac{\gamma D_R}{2 k}\right)\bm{1}-\frac{\gamma \omega}{2 k} \underline{\underline{U}} \right] \left( D_R\bm{1}-\omega \underline{\underline{U}}\right)\right\}^{-1}=\\
    &=\frac{1}{\left[ D_R \left(1+\frac{\gamma D_R}{2k} \right)-\frac{\omega^2\gamma}{2k} \right]^2+\left[ \omega \left(1+\frac{\gamma D_R}{2k} \right)+\frac{D_R \omega \gamma}{2k} \right]^2} \begin{pmatrix}D_R \left(1+\frac{\gamma D_R}{2k} \right)-\frac{\omega^2\gamma}{2k} & -\omega \left(1+\frac{\gamma D_R}{2k} \right)-\frac{D_R \omega \gamma}{2k} \\
    \omega \left(1+\frac{\gamma D_R}{2k} \right)+\frac{D_R \omega \gamma}{2k} & D_R \left(1+\frac{\gamma D_R}{2k} \right)-\frac{\omega^2\gamma}{2k}\end{pmatrix}
    \end{split}
\end{equation}
As in \eqref{matrix_1}, the anti-diagonal terms cancel when we consider the sum
\begin{equation}
\begin{split}
    &\left( D_R\bm{1}-\omega \underline{\underline{U}}\right)^{-1} \left[ \left( 1+\frac{\gamma D_R}{2 k}\right)\bm{1}-\frac{\gamma \omega}{2 k} \underline{\underline{U}} \right]^{-1} + \left( D_R\bm{1}+\omega \underline{\underline{U}}\right)^{-1} \left[ \left( 1+\frac{\gamma D_R}{2 k}\right)\bm{1}+\frac{\gamma \omega}{2 k} \underline{\underline{U}} \right]^{-1}=\\
    &=\frac{2\left[ D_R \left(1+\frac{\gamma D_R}{2k} \right)-\frac{\omega^2\gamma}{2k} \right]}{\left[ D_R \left(1+\frac{\gamma D_R}{2k} \right)-\frac{\omega^2\gamma}{2k} \right]^2+\left[ \omega \left(1+\frac{\gamma D_R}{2k} \right)+\frac{D_R \omega \gamma}{2k} \right]^2} \bm{1}
    \end{split}
\end{equation}
Thus, the total flux reads:
\begin{equation}
    \begin{split}
        \bm{J}&=-\frac{1}{16} \frac{2\left[ D_R \left(1+\frac{\gamma D_R}{2k} \right)-\frac{\omega^2\gamma}{2k} \right]}{\left[ D_R \left(1+\frac{\gamma D_R}{2k} \right)-\frac{\omega^2\gamma}{2k} \right]^2+\left[ \omega \left(1+\frac{\gamma D_R}{2k} \right)+\frac{D_R \omega \gamma}{2k} \right]^2}  \left(\nabla_{\bm{R}} v^2_a \left(\bm{R}\right)\right) \rho(\bm{R})+\\
        &-\left[ \frac{1}{8} \frac{2 D_R}{D^2_R+\omega^2} v^2_a \left(\bm{R}\right) +\frac{D_T}{2} \right] \nabla_{\bm{R}} \rho(\bm{R})
    \end{split}
\end{equation}
By comparing this equation with \eqref{coarse_grained_FP} we can identify the drift and the diffusion terms with:
\begin{equation}
    \begin{cases}
    \bm{V}(\bm{R})=-\frac{1}{8} \frac{\big[D_R\big( 1+\frac{\gamma D_R}{2k}\big) - \frac{\omega^2 \gamma}{2k}\big]\nabla_{\bm{R}} v_a^2(\bm{R})}{\big[D_R\big( 1+\frac{\gamma D_R}{2k}\big) - \frac{\omega^2 \gamma}{2k}\big]^2 + \big[ \omega \big(1 + \frac{\gamma D_R}{2k} \big) + \frac{D_R \omega \gamma}{2k}  \big]^2} \\
    \\
    D(\bm{R})=\frac{1}{4}\frac{D_R}{D_R^2 + \omega^2} v_a^2(\bm{R}) + \frac{D_T}{2}
    \end{cases}
\end{equation}
Moreover, the drift and diffusion terms can be related as follows:
\begin{equation}
    \begin{cases}
    \bm{V}(\bm{R}) = -\frac{1}{2} \epsilon \nabla_{R} D(\bm{R}) \\
    \\
    \epsilon=\frac{\big[D_R\big( 1+\frac{\gamma D_R}{2k}\big) - \frac{\omega^2 \gamma}{2k}\big]\frac{D_R^2 + \omega^2}{D_R}}{\big[D_R\big( 1+\frac{\gamma D_R}{2k}\big) - \frac{\omega^2 \gamma}{2k}\big]^2 + \big[ \omega \big(1 + \frac{\gamma D_R}{2k} \big) + \frac{D_R \omega \gamma}{2k}  \big]^2}
    \end{cases}
\end{equation}
This allows us to find the steady state density without specifying a particular form for the activity function. In particular, imposing $\bm{J}=0$ at steady state, we get:
\begin{equation}
    \rho(\mathbf{R}) \propto \Big[ \frac{1}{4}\frac{D_R}{D_R^2 + \omega^2} v_a^2(\mathbf{R}) + \frac{D_T}{2} \Big]^{-\frac{1}{2}\epsilon}
\end{equation}
\paragraph{Case of rigid bond}
The previous derivation can be specialized to the case of a rigid bond by considering the harmonic potential with finite rest length $V(\bm{r})=\frac{k}{2}\left(r-l_0 \right)^2$ and later taking the limit $k \to \infty$. In particular, we have that $\nabla \bm{F}=-k\underline{\underline{A}}$ with $\underline{\underline{A}}=\hat{\bm{r}}\hat{\bm{r}}+\left( \bm{1-\hat{\bm{r}}\hat{\bm{r}}}\right)\left( 1-\frac{l_0}{r} \right)$. Being $\underline{\underline{A}}$ a symmetric matrix, Equation \eqref{I_1} can be rewritten as:
\begin{equation}
    \begin{split}
        \bm{I}_1&=\int d\bm{r}\, \frac{1}{4 \gamma} \left[ \nabla_{\bm{R}}  v_a \Big(\bm{R}+\frac{\bm{r}}{2} \Big)\cdot \bm{F}\right] \bm{\sigma}_{1}=\\&=+\frac{1}{4 \gamma}\left( D_R\bm{1}-\omega \underline{\underline{U}} \right)^{-1} \int d\bm{r}\,  \left[ \nabla_{\bm{R}}  v_a \Big(\bm{R}+\frac{\bm{r}}{2} \Big) \cdot \left(\nabla_{\bm{r}} \bm{F} \right)\right] \cdot \Big[ \frac{2}{\gamma} \bm{F}\bm{\sigma}_{1} + v_a \Big(\bm{R}+\frac{\bm{r}}{2} \Big)\phi \bm{1}\Big]=\\
        &=\left( D_R\bm{1}-\omega \underline{\underline{U}} \right)^{-1} \left[-\frac{2k}{\gamma} \bm{I}_1 -\frac{k}{4 \gamma} \int d\bm{r}\,   \nabla_{\bm{R}}  v_a \Big(\bm{R}+\frac{\bm{r}}{2} \Big) \cdot \underline{\underline{A}} \, v_a \Big(\bm{R}+\frac{\bm{r}}{2} \Big)\phi \right]=\\
        &=-\frac{1}{8}\left[ \left( 1+\frac{\gamma D_R}{2 k}\right)\bm{1}-\frac{\gamma \omega}{2 k} \underline{\underline{U}} \right]^{-1} \int d\bm{r}\,v_a \Big(\bm{R}+\frac{\bm{r}}{2} \Big) \phi \nabla_{\bm{R}}  v_a \Big(\bm{R}+\frac{\bm{r}}{2} \Big) \cdot \underline{\underline{A}}
    \end{split}
\end{equation}
from which we can compute the flux
\begin{equation}
\begin{split}
        \bm{J}_{
\bm{\sigma}_1}&=\left( D_R\bm{1}-\omega \underline{\underline{U}}    \right)^{-1} \Big\{-\int d\bm{r}\, \frac{1}{8}v_a \left(\bm{R}\right) \nabla_{\bm{R}} \cdot \Big[ v_a \left(\bm{R}\right)\phi \bm{1}\Big] + \int d\bm{r}\, \frac{1}{8} \left[ \nabla_{\bm{R}}  v_a \left(\bm{R}\right)\right] v_a \left(\bm{R}\right)\phi + \\
& -\frac{1}{8}\left[ \left( 1+\frac{\gamma D_R}{2 k}\right)\bm{1}-\frac{\gamma \omega}{2 k} \underline{\underline{U}} \right]^{-1} \int d\bm{r}\,v_a \left(\bm{R}\right) \phi \nabla_{\bm{R}}  v_a \left(\bm{R}\right) \cdot \underline{\underline{A}}\Big\}
\end{split}
\end{equation}
In the last expression, the dependence of $v_a\left(\bm{R}+\frac{\bm{r}}{2} \right)$ on the inner coordinate has been dropped as it leads to corrections proportional to $\mathcal{O}(\nabla^2)$, which are negligible in the limit of rest length much smaller than the activity gradient. Moreover, for $k \to \infty$, we can approximate $\phi(\bm{R},\bm{r},t) \approx \rho(\bm{R},t)\delta\left( r-l_0\right)\frac{1}{2 \pi l_0}$ and
\begin{equation}
    \lim_{k \to \infty}\left[ \left( 1+\frac{\gamma D_R}{2 k}\right)\bm{1}-\frac{\gamma \omega}{2 k} \underline{\underline{U}} \right]^{-1} = \bm{1}
\end{equation}
In this way, the flux becomes:
\begin{equation}
\begin{split}
        \bm{J}_{
\bm{\sigma}_1}&=\left( D_R\bm{1}-\omega \underline{\underline{U}}    \right)^{-1} \Big\{- \frac{1}{8}v_a^2 \left(\bm{R}\right) \nabla_{\bm{R}}  \rho(\bm{R}) -\frac{1}{16} \rho(\bm{R}) \nabla_{\bm{R}}  v_a^2 \left(\bm{R}\right) \cdot \int d\bm{r}\, \frac{1}{2 \pi l_0}\delta (r-l_0)\underline{\underline{A}} \Big\}=\\
&=\left( D_R\bm{1}-\omega \underline{\underline{U}}    \right)^{-1} \Big\{- \frac{1}{8}v_a^2 \left(\bm{R}\right) \nabla_{\bm{R}}  \rho(\bm{R}) -\frac{1}{32} \rho(\bm{R}) \nabla_{\bm{R}}  v_a^2 \left(\bm{R}\right) \Big\}
\end{split}
\end{equation}
Analogously:
\begin{equation}
    \bm{J}_{
\bm{\sigma}_2}=\left( D_R\bm{1}+\omega \underline{\underline{U}}    \right)^{-1} \Big\{- \frac{1}{8}v_a^2 \left(\bm{R}\right) \nabla_{\bm{R}}  \rho(\bm{R}) -\frac{1}{32} \rho(\bm{R}) \nabla_{\bm{R}}  v_a^2 \left(\bm{R}\right) \Big\}
\end{equation}
Thus, the total flux reads:
\begin{equation}
\begin{split}
    \bm{J}&=\frac{2D_R}{D^2_R+\omega^2} \left[ - \frac{1}{8}v_a^2 \left(\bm{R}\right) \nabla_{\bm{R}}  \rho(\bm{R}) -\frac{1}{32} \rho(\bm{R}) \nabla_{\bm{R}}  v_a^2 \left(\bm{R}\right) \right] -\frac{D_T}{2}\nabla_{\bm{R}}\rho(\bm{R})=\\
    &=-\left(\frac{1}{4}\frac{D_R}{D_R^2+\omega^2 }v_a^2(\bm{R}) + \frac{D_T}{2} \right) \nabla_{\bm{R}} \rho(\bm{R}) - \frac{1}{16} \frac{D_R}{D_R^2+\omega^2 } \nabla_{\bm{R}}v_a^2(\bm{R}) \rho(\bm{R})
\end{split}
\end{equation}
Imposing zero flux at steady state, we find:
\begin{equation}
    \rho(\bm{R})\propto \left( \frac{1}{4}\frac{D_R}{D_R^2+\omega^2 }v_a^2(\bm{R}) + \frac{D_T}{2}  \right)^{-\frac{1}{4}}
\end{equation}

\section{Single Active Chiral Particle}
\label{sec_SACP}
In this section we focus on a single active chiral particle described by the dynamics
\begin{equation}
    \begin{split}
    &\frac{d\bm{r}}{dt}=v_a(\bm{r})\bm{p}(\theta) + \sqrt{2 D_T} \bm{\xi}(t)\\
    &\frac{d \theta}{dt}=\omega(\bm{r}) + \sqrt{2 D_R} \eta(t)
    \end{split}
\end{equation}
in the general case where both the activity function $v_a(\bm{r})$ and the active torque $\omega(\bm{r})$ are slowly varying non-homogeneous functions of space. The self propulsion force is directed along the orientation vector $\bm{p}(\theta)=\left(\cos \theta, \sin \theta \right)$, and $\bm{\xi}(t)$, $\eta(t)$ are independent gaussian white noises. The probability density $P(\bm{r},\theta,t)$ evolves according to the following FPE:
\begin{equation}
    \partial_t P(\bm{r},\theta,t)= -\nabla \cdot \left[ v_a(\bm{R}) \bm{p}(\theta)P-D_T \nabla P \right] - \partial_\theta \left[ \omega(\bm{r})P \right] + D_R\partial^2_\theta P
\end{equation}
To integrate out the rotational degree of freedom we expand the probability density using the eigenfunctions of the operator $\partial^2_\theta$:
\begin{equation}
    P(\bm{r},\theta,t)=\frac{1}{\Omega_2}\left[ \rho(\bm{r},t) + \bm{\sigma}(\bm{r},t)\cdot \bm{p} + \underline{\underline{\bm{w}}}(\bm{r},t):\left( \bm{p}\bm{p} - \frac{\bm{1}}{2}\right) + h.o.t. \right]
\label{proj_sacp}
\end{equation}
and then we project Equation \eqref{proj_sacp} onto such eigenfunctions. This procedure leads to a hierarchy of equations for the coefficients of the expansion:
\begin{equation}
    \begin{split}
    &\partial_t \rho(\bm{r},t)=-\nabla \cdot \left[ \frac{1}{2}v_a(\bm{r})\bm{\sigma} - D_T \nabla \rho\right]\\
    &\partial_t \bm{\sigma}(\bm{r},t)=-\nabla \left[ v_a(\bm{r}) \rho \right] -\frac{1}{2} \nabla \cdot \left[ v_a(\bm{r}) \underline{\underline{\bm{w}}} \right] + D_T \nabla^2 \bm{\sigma} - \left( D_R \bm{1} - \omega(\bm{r})\underline{\underline{U}}\right) \bm{\sigma}\\
    &...
    \end{split}
\end{equation}
with the anti-symmetric matrix $\underline{\underline{U}}$ already defined in the previous section and "..." standing for all the other equations in the hierarchy. Considering that $\bm{\sigma}(\bm{r},t)$ relaxes on a much smaller time scale than $\rho(\bm{r},t)$ and neglecting all terms proportional to $\mathcal{O}(\nabla^2)$ or higher order derivatives (small gradient approximation), we get:
\begin{equation}
\begin{split}
    &\partial_t \rho(\bm{r},t)=-\nabla \cdot \left[ \frac{1}{2}v_a(\bm{r})\bm{\sigma} - D_T \nabla \rho\right]\\
    &\left( D_R \bm{1} - \omega(\bm{r})\underline{\underline{U}}\right) \bm{\sigma}=-\nabla \left[ v_a(\bm{r}) \rho \right]  \\
    \end{split}
\end{equation}
After defining
\begin{equation}
    \underline{\underline{M}}=\left(\bm{1} - \frac{\omega(\bm{r})}{D_R}\underline{\underline{U}}\right)
\end{equation}
and plugging the expression of $\bm{\sigma}$ in the first equation, we obtain the effective FPE for $\rho(\bm{r},t)$
\begin{equation}
\partial_t \rho(\bm{r})= -\nabla_{\bm{r}} \cdot \Big[ -\frac{v_a(\bm{r})}{2D_R}  \underline{\underline{\bm{M}}}^{-1} \nabla v_a(\bm{r})\rho(\bm{r}) - \left( \frac{v^2_a(\bm{r})}{2D_R} \underline{\underline{\bm{M}}}^{-1} + D_T\bm{1}\right) \nabla_{\bm{r}}\rho(\bm{r}) \Big]
\end{equation}
characterized by the following drift and diffusion terms:
\begin{equation}
\begin{split}
        &\bm{V}^C(\bm{r})=-\frac{v_a(\bm{r})}{2D_R}  \underline{\underline{\bm{M}}}^{-1} \nabla v_a(\bm{r}),\\
    \\
    &\underline{\underline{\bm{D}}}^C(\bm{r})=\frac{v^2_a(\bm{r})}{2D_R} \underline{\underline{\bm{M}}}^{-1} + D_T\bm{1},\\
    \end{split}
\end{equation}

\section{Single Active Particle under Lorentz force}
\label{sec_SAPB}
An analogous derivation can be done for a single active charged particle immersed in a magnetic field. The magnetic field is in general non-homogeneous in space and its effect can be included in the effective friction tensor
\begin{equation}
    \gamma \underline{\underline{\Gamma}} = \gamma \begin{pmatrix}
        1 & -\kappa(\bm{r}) \\
        \kappa(\bm{r}) & 1 
        \end{pmatrix}
\end{equation}
characterized by anti-symmetric components proportional to $\kappa(\bm{r})=\frac{qB(\bm{r})}{\gamma}$. 
As in the previous section, we start our analysis from the FPE in the small mass limit ~\cite{vuijk2020lorentz,abdoli2021stochastic}:
\begin{equation}
    \partial_t P(\bm{r},\theta,t)=-\nabla \cdot \Big[ \underline{\underline{\Gamma}}^{-1}(\bm{r}) \nu_a(\bm{r}) \bm{p}P - D_T \underline{\underline{\Gamma}}^{-1}(\bm{r}) \nabla P \Big] + D_R \frac{\partial^2}{\partial \theta^2}P,
    \label{fpe_samp}
\end{equation}
and we derive an evolution equation for the density $\rho(\bm{r},t)=\int d\theta \,P(\bm{r},\theta,t)$ under the assumption of slowly varying activity function. Using again the decomposition in \eqref{proj_sacp} and projecting \eqref{fpe_samp} onto the eigenfunctions of $\partial_\theta^2$, we obtain the following equations:
\begin{equation}
    \begin{split}
        &\partial_t \rho(\bm{r},t)=-\nabla \cdot \left[ \frac{\nu_a(\bm{r})}{2} \underline{\underline{\Gamma}}^{-1}  \bm{\sigma}(\bm{r}) - D_T \underline{\underline{\Gamma}}^{-1} \nabla \rho(\bm{r}) \right]\\
        \\
        &\partial_t \bm{\sigma}(\bm{r},t)=-\nabla \cdot \left[ \nu_a(\bm{r})\rho(\bm{r}) \underline{\underline{\Gamma}}^{-1} -D_T \underline{\underline{\Gamma}}^{-1} \nabla \bm{\sigma}(\bm{r}) \right] - D_R \bm{\sigma}(\bm{r})
    \end{split}
\end{equation}
where all terms proportional to $\underline{\underline{w}}$ have already been neglected, as well as all the other equations of the hierarchy. The equation for the average polarization can be further simplified considering that $\bm{\sigma}(\bm{r})$ relaxes on a much smaller time scale than $\rho(\bm{r})$ and that spatial gradients in the system are small. In particular
\begin{equation}
    \bm{\sigma}(\bm{r})=-\nabla \cdot \left[ \frac{1}{D_R} \nu_a(\bm{r})\rho(\bm{r}) \underline{\underline{\Gamma}}^{-1}\right]
\end{equation}
and the evolution of $\rho(\bm{r},t)$ reads:
\begin{equation}
    \partial_t \rho(\bm{r})=-\nabla \cdot \left[ -\frac{1}{2D_R} \underline{\underline{\Gamma}}^{-1} \nu_a(\bm{r}) \nabla \cdot \left( \underline{\underline{\Gamma}}^{-1} \nu_a(\bm{r}) \right) \rho(\bm{r}) - \left( \frac{1}{2D_R} \nu^2_a(\bm{r}) \underline{\underline{\Gamma}}^{-1} \cdot \left(\underline{\underline{\Gamma}}^{-1}\right)^T + D_T \underline{\underline{\Gamma}}^{-1} \right) \nabla \rho(\bm{r})\right]
\end{equation}
Being $\underline{\underline{\Gamma}}^{-1} \cdot \left(\underline{\underline{\Gamma}}^{-1}\right)^T=\left( 1+\kappa^2(\bm{r})\right)^{-1} \bm{1}$, the drift and diffusion terms can be identified with
\begin{equation}
\begin{split}
    &\bm{V}^B(\bm{r})=-\frac{\nu_a({\bm{r}})}{2 D_R} \underline{\underline{\bm{\Gamma}}}^{-1} \nabla \cdot \left( \nu_a({\bm{r}}) \underline{\underline{\bm{\Gamma}}}^{-1} \right)\\
    \\
    &\underline{\underline{\bm{D}}}^B(\bm{r})=\frac{1}{1+\kappa^2(\bm{r})}\begin{pmatrix}
    \frac{\nu^2_a(\bm{r})}{2 D_R} + D_T  & D_T \kappa(\bm{r}) \\
    - D_T \kappa(\bm{r}) & \frac{\nu^2_a(\bm{r})}{2 D_R} + D_T 
    \end{pmatrix}
    \end{split}
\end{equation}

\section{Active dimer under Lorentz force}
\label{sec_dimerB}
In this section we focus on a dimer composed of two interacting active charged particles (with opposite charges $\pm q$) immersed in a homogeneous magnetic field $\bm{B}$. Also in this case we restrict our analysis to dimension $d=2$. In the following the interaction potential is modeled as a spring with stiffness $k_s$ and zero rest length, but the derivation can be easily generalized for the case of finite rest length and rigid bond interaction. The probability $P(\bm{r}_1,\bm{r}_2,\theta_1, \theta_2,t)$ to find the system in a certain configuration at time $t$ evolves according to the following FPE:
\begin{equation}
    \begin{split}
        \partial_t P(\bm{r}_1,\bm{r}_2,\theta_1, \theta_2,t)=& -\nabla_{\bm{r}_1} \cdot \left[ \frac{1}{\gamma}\underline{\underline{\Gamma}}_+^{-1} \bm{F} P + \underline{\underline{\Gamma}}_+^{-1} v_a(\bm{r}_1) \bm{p}_1 P - D_T \underline{\underline{\Gamma}}_+^{-1} \nabla_{\bm{r}_1}P\right] + \\
        &-\nabla_{\bm{r}_2} \cdot \left[ -\frac{1}{\gamma}\underline{\underline{\Gamma}}_-^{-1} \bm{F} P + \underline{\underline{\Gamma}}_-^{-1} v_a(\bm{r}_2) \bm{p}_2 P - D_T \underline{\underline{\Gamma}}_-^{-1} \nabla_{\bm{r}_2}P\right]+D_R \left[ \partial_{\theta_1}^2+\partial_{\theta_2}^2 \right]P
    \end{split}
    \label{fpe_dimerB}
\end{equation}
where
\begin{equation}
    \underline{\underline{\Gamma}}_{\pm}=\frac{1}{1+\kappa^2}\begin{pmatrix} 1 & \mp \kappa\\
    \pm \kappa & 1
    \end{pmatrix},
\end{equation}
$\kappa=\frac{+ q B}{\gamma}$, $\bm{p}_i=\left( \cos \theta_i, \sin \theta_i\right)$ are the orientation vectors setting the direction of the active self propulsion forces and $\bm{F}=-\nabla_{\bm{r}_1} U(\bm{r}_1-\bm{r}_2)$. The parameters $D_T$ and $D_R$ are related respectively to translational and rotational diffusion. We find convenient to express Equation \eqref{fpe_dimerB} in terms of the new set of coordinates
\begin{equation}
    \begin{cases}
        \bm{R}=\left[ \gamma \left( \underline{\underline{\Gamma}}_+ + \underline{\underline{\Gamma}}_- \right) \right]^{-1} \left[\gamma \underline{\underline{\Gamma}}_+ \bm{r}_1 + \gamma \underline{\underline{\Gamma}}_- \bm{r}_2 \right]=\frac{1}{2} \left[ \underline{\underline{\Gamma}}_+ \bm{r}_1 + \underline{\underline{\Gamma}}_- \bm{r}_2 \right]\\
        \\
        \bm{r}=\bm{r}_1-\bm{r}_2
    \end{cases}
\end{equation}
that we call center-of-friction and inner coordinate. In particular, the new FPE reads:
\begin{equation}
\begin{split}
    \partial_t P(\bm{R},\bm{r},\theta_1, \theta_2,t)=&-\nabla_{\bm{R}} \cdot \left[ \frac{1}{2} v_a(\bm{r}_1) \bm{p}_1 P + \frac{1}{2} v_a(\bm{r}_2) \bm{p}_2 P - \frac{D_T}{2} \nabla_{\bm{R}}P\right] +\\
    &-\nabla_{\bm{r}} \cdot \left[ \frac{1}{\gamma} \frac{2}{1+\kappa^2} \bm{F} P + \underline{\underline{\Gamma}}_+^{-1} v_a(\bm{r}_1) \bm{p}_1 P  - \underline{\underline{\Gamma}}_-^{-1} v_a(\bm{r}_2) \bm{p}_2 P - \frac{2 D_T}{1 + \kappa^2} \nabla_{\bm{r}}P \right] + \\
    &+\frac{D_T}{2}\frac{4\kappa}{1+\kappa^2} \left(\partial_{r_x}\partial_{R_y} - \partial_{r_y}\partial_{R_x} \right) P +D_R \left[ \partial_{\theta_1}^2+\partial_{\theta_2}^2 \right]P
\end{split}
\label{fpe_dimerB_2}
\end{equation}
with $\bm{r}_1=\bm{R} + \frac{1}{2}\underline{\underline{\Gamma}}_- \bm{r}$ and $\bm{r}_2=\bm{R} - \frac{1}{2}\underline{\underline{\Gamma}}_+ \bm{r}$. At this point, the strategy to integrate out the angular degrees of freedom is analogous to the one presented in the first section, namely it relies on the expansion:
\begin{equation}
    P(\bm{R},\bm{r},\theta_1,\theta_2,t)=\frac{1}{\Omega_2^2} \left[\phi + \bm{\sigma}_1 \cdot \bm{p}_1 + \bm{\sigma}_2 \cdot \bm{p}_2 + \underline{\underline{\bm{\sigma}}}_{12} : \bm{p}_1\bm{p}_2 +  \underline{\underline{\bm{w}}}_1 :\left( \bm{p}_1\bm{p}_1 -\frac{1}{2}\bm{I} \right) +  \underline{\underline{\bm{w}}}_2 :\left( \bm{p}_2\bm{p}_2 -\frac{1}{2}\bm{I} \right) + \Theta\right]
\end{equation}
with $\Theta$ indicating higher order terms and $\Omega_2=2\pi$. However, we will neglect contributions coming from $\underline{\underline{\bm{\sigma}}}_{12},\underline{\underline{\bm{w}}}_1 , \underline{\underline{\bm{w}}}_2, \Theta$ as, similarly to Section $1$, they lead to corrections of order $\mathcal{O}(\nabla_{\bm{R}}^2)$ or higher and we are working in the approximation of slowly varying activity. Thus, projecting Equation \eqref{fpe_dimerB_2} onto the eigenfunctions $1,\bm{p}_1,\bm{p}_2$, we get the following set of equations for the coefficients $\phi,\bm{\sigma}_1, \bm{\sigma}_2$:
\begin{equation}
\begin{split}
    \partial_t \phi(\bm{R},\bm{r},t)=&-\nabla_{\bm{R}} \cdot \left[ \frac{1}{2} v_a(\bm{r}_1) \frac{\bm{\sigma}_1}{2} + \frac{1}{2} v_a(\bm{r}_2) \frac{\bm{\sigma}_2}{2} - \frac{D_T}{2} \nabla_{\bm{R}} \phi \right] + \\
    &-\nabla_{\bm{r}} \cdot \left[ \frac{1}{\gamma} \frac{2}{1+\kappa^2} \bm{F} \phi + \underline{\underline{\Gamma}}_+^{-1} v_a(\bm{r}_1) \frac{\bm{\sigma}_1}{2}  - \underline{\underline{\Gamma}}_-^{-1} v_a(\bm{r}_2) \frac{\bm{\sigma_2}}{2} - \frac{2 D_T}{1 + \kappa^2} \nabla_{\bm{r}} \phi \right]+\\
    &+\frac{D_T}{2}\frac{4\kappa}{1+\kappa^2} \left(\partial_{r_x}\partial_{R_y} - \partial_{r_y}\partial_{R_x} \right) \phi,
\end{split}
\label{phi_dimer_B}
\end{equation}
\begin{equation}
    \begin{split}
        \partial_t \bm{\sigma}_1=&-\nabla_{\bm{R}} \cdot \left[ \frac{1}{2} v_a(\bm{r}_1)\phi \bm{1} - \frac{D_T}{2} \nabla_{\bm{R}} \bm{\sigma}_1 \right] -\nabla_{\bm{r}} \cdot \left[ \frac{1}{\gamma} \frac{2}{1+\kappa^2} \bm{F} \bm{\sigma}_1 + v_a(\bm{r}_1)\phi \underline{\underline{\Gamma}}_+^{-1} - \frac{2 D_T}{1+\kappa^2} \nabla_{\bm{r}} \bm{\sigma}_1 \right] + \\
        &+\frac{D_T}{2}\frac{4\kappa}{1+\kappa^2} \left(\partial_{r_x}\partial_{R_y} - \partial_{r_y}\partial_{R_x} \right) \bm{\sigma}_1 -D_R \bm{\sigma}_1,
    \end{split}
    \label{sigma1_dimer_B}
\end{equation}
and
\begin{equation}
    \begin{split}
        \partial_t \bm{\sigma}_2=&-\nabla_{\bm{R}} \cdot \left[ \frac{1}{2} v_a(\bm{r}_2)\phi \bm{1} - \frac{D_T}{2} \nabla_{\bm{R}} \bm{\sigma}_2 \right] -\nabla_{\bm{r}} \cdot \left[ \frac{1}{\gamma} \frac{2}{1+\kappa^2} \bm{F} \bm{\sigma}_2 - v_a(\bm{r}_2)\phi \underline{\underline{\Gamma}}_-^{-1} - \frac{2 D_T}{1+\kappa^2} \nabla_{\bm{r}} \bm{\sigma}_2 \right] + \\
        &+\frac{D_T}{2}\frac{4\kappa}{1+\kappa^2} \left(\partial_{r_x}\partial_{R_y} - \partial_{r_y}\partial_{R_x} \right) \bm{\sigma}_2 -D_R \bm{\sigma}_2,
    \end{split}
    \label{sigma2_dimer_B}
\end{equation}
The evolution of the density $\rho(\bm{R},t)=\int \, d \bm{r} \phi(\bm{R},\bm{r},t)$ can be obtained by integrating Equation \eqref{phi_dimer_B} over the inner coordinate:
\begin{equation}
    \partial_t \rho(\bm{R},t)=-\nabla_{\bm{R}} \cdot \bm{J}=-\nabla_{\bm{R}} \cdot \left[ \int d \bm{r} \left( \frac{1}{4} v_a(\bm{r}_1) \bm{\sigma}_1 + \frac{1}{4} v_a(\bm{r}_2) \bm{\sigma}_2 \right) - \frac{D_T}{2} \nabla_{\bm{R}} \rho \right]
    \label{coarse_grained_fpe_dimerB}
\end{equation}
The flux $\bm{J}$ can be divided into three main contributions
\begin{equation}
    \begin{cases}
        \bm{J}_{\bm{\sigma}_1}=\int d \bm{r} \frac{1}{4}v_a\left(\bm{R} + \frac{1}{2}\underline{\underline{\Gamma}}_- \bm{r}\right) \bm{\sigma}_1\\
        \\
        \bm{J}_{\bm{\sigma}_2}=\int d \bm{r} \frac{1}{4}v_a\left(\bm{R} - \frac{1}{2}\underline{\underline{\Gamma}}_+ \bm{r}\right) \bm{\sigma}_2\\
        \\
        \bm{J}_{D}=\frac{D_T}{2}\nabla_{\bm{R}} \rho
    \end{cases}
\end{equation}
Since the average polarizations $\bm{\sigma}_1$ and $\bm{\sigma}_2$ relax on a much smaller time scale than $\rho(\bm{R},t)$, we can simplify the equations \eqref{sigma1_dimer_B} and \eqref{sigma2_dimer_B} as:
\begin{equation}
    \begin{split}
        \bm{\sigma}_1=&-\frac{1}{D_R}\nabla_{\bm{R}} \cdot \left[ \frac{1}{2} v_a(\bm{r}_1)\phi \bm{1} - \frac{D_T}{2} \nabla_{\bm{R}} \bm{\sigma}_1 \right] -\frac{1}{D_R}\nabla_{\bm{r}} \cdot \left[ \frac{1}{\gamma} \frac{2}{1+\kappa^2} \bm{F} \bm{\sigma}_1 + v_a(\bm{r}_1)\phi \underline{\underline{\Gamma}}_+^{-1} - \frac{2 D_T}{1+\kappa^2} \nabla_{\bm{r}} \bm{\sigma}_1 \right] + \\
        &+\frac{1}{D_R}\frac{D_T}{2}\frac{4\kappa}{1+\kappa^2} \left(\partial_{r_x}\partial_{R_y} - \partial_{r_y}\partial_{R_x} \right) \bm{\sigma}_1,
    \end{split}
    \label{sigma1_dimerB}
\end{equation}
\begin{equation}
    \begin{split}
         \bm{\sigma}_2=&-\frac{1}{D_R}\nabla_{\bm{R}} \cdot \left[ \frac{1}{2} v_a(\bm{r}_2)\phi \bm{1} - \frac{D_T}{2} \nabla_{\bm{R}} \bm{\sigma}_2 \right] -\frac{1}{D_R}\nabla_{\bm{r}} \cdot \left[ \frac{1}{\gamma} \frac{2}{1+\kappa^2} \bm{F} \bm{\sigma}_2 - v_a(\bm{r}_2)\phi \underline{\underline{\Gamma}}_-^{-1} - \frac{2 D_T}{1+\kappa^2} \nabla_{\bm{r}} \bm{\sigma}_2 \right] + \\
        &+\frac{1}{D_R}\frac{D_T}{2}\frac{4\kappa}{1+\kappa^2} \left(\partial_{r_x}\partial_{R_y} - \partial_{r_y}\partial_{R_x} \right) \bm{\sigma}_2
    \end{split}
\end{equation}
In the following we focus on the flux contribution $\bm{J}_{\bm{\sigma}_1}$: 
\begin{equation}
\begin{split}
    \bm{J}_{\bm{\sigma}_1}&=\int d \bm{r} \frac{1}{4}v_a\left(\bm{R} + \frac{1}{2}\underline{\underline{\Gamma}}_- \bm{r}\right) \bm{\sigma}_1=\\
    &=\int d \bm{r} \frac{1}{4D_R}v_a\left(\bm{R} + \frac{1}{2}\underline{\underline{\Gamma}}_- \bm{r}\right) \Bigg\{ -\nabla_{\bm{R}} \cdot \left[ \frac{1}{2} v_a(\bm{r}_1)\phi \bm{1} - \frac{D_T}{2} \nabla_{\bm{R}} \bm{\sigma}_1 \right]+\\ &-\nabla_{\bm{r}} \cdot \left[ \frac{1}{\gamma} \frac{2}{1+\kappa^2} \bm{F} \bm{\sigma}_1 + v_a(\bm{r}_1)\phi \underline{\underline{\Gamma}}_+^{-1} - \frac{2 D_T}{1+\kappa^2} \nabla_{\bm{r}} \bm{\sigma}_1 \right] + \frac{D_T}{2}\frac{4\kappa}{1+\kappa^2} \left(\partial_{r_x}\partial_{R_y} - \partial_{r_y}\partial_{R_x} \right) \bm{\sigma}_1\Bigg\}
\end{split}
\end{equation}
Note that gradients $\nabla_{\bm{r}}$ are in general not small, but they can be turned into gradients with respect to $\bm{R}$ integrating by part and using 
\begin{equation}
    \nabla_{\bm{r}} v_a\left(\bm{R} + \frac{1}{2}\underline{\underline{\Gamma}}_- \bm{r}\right)=\frac{1}{2} \underline{\underline{\Gamma}}_-^T \cdot \nabla_{\bm{R}} v_a\left(\bm{R} + \frac{1}{2}\underline{\underline{\Gamma}}_- \bm{r}\right)
\end{equation}
Then, all terms proportional to $\mathcal{O}(\nabla_{\bm{R}}^2)$ can be neglected as we are considering swim forces which vary slowly in space. This approximation leads to
\begin{equation}
    \begin{split}
        \bm{J}_{\bm{\sigma}_1}&=\int d \bm{r} \frac{1}{4D_R}v_a\left(\bm{R} + \frac{1}{2}\underline{\underline{\Gamma}}_- \bm{r}\right) \Bigg\{ -\nabla_{\bm{R}} \left[ \frac{1}{2} v_a(\bm{r}_1)\phi \right]-\nabla_{\bm{r}} \cdot \left[ \frac{1}{\gamma} \frac{2}{1+\kappa^2} \bm{F} \bm{\sigma}_1 + v_a(\bm{r}_1)\phi \underline{\underline{\Gamma}}_+^{-1} \right] \Bigg\}
    \end{split}
\end{equation}
The previous expression can be manipulated to get:
\begin{equation}
\begin{split}
    \bm{J}_{\bm{\sigma}_1}=&-\frac{1}{8D_R}\int d \bm{r}\, v_a\left(\bm{R} + \frac{1}{2}\underline{\underline{\Gamma}}_- \bm{r}\right) \nabla_{\bm{R}} \left[ v_a\left(\bm{R} + \frac{1}{2}\underline{\underline{\Gamma}}_- \bm{r}\right) \phi(\bm{R},\bm{r}) \right]+\\
    &-\frac{1}{4D_R}\int d \bm{r}\, v_a\left(\bm{R} + \frac{1}{2}\underline{\underline{\Gamma}}_- \bm{r}\right) \nabla_{\bm{r}} \cdot \left[ v_a\left(\bm{R} + \frac{1}{2}\underline{\underline{\Gamma}}_- \bm{r}\right) \phi(\bm{R},\bm{r}) \underline{\underline{\Gamma}}_+^{-1} + \frac{1}{\gamma} \frac{2}{1+\kappa^2} \bm{F} \bm{\sigma}_1 \right] + \\
    &=-\frac{1}{8D_R}\int d \bm{r}\, v_a\left(\bm{R} + \frac{1}{2}\underline{\underline{\Gamma}}_- \bm{r}\right) \nabla_{\bm{R}} \left[ v_a\left(\bm{R} + \frac{1}{2}\underline{\underline{\Gamma}}_- \bm{r}\right) \phi(\bm{R},\bm{r}) \right]+\\
    &+\frac{1}{4D_R}\int d \bm{r}\, \left(\nabla_{\bm{r}} v_a\left(\bm{R} + \frac{1}{2}\underline{\underline{\Gamma}}_- \bm{r}\right)\right)  \cdot \left[ v_a\left(\bm{R} + \frac{1}{2}\underline{\underline{\Gamma}}_- \bm{r}\right) \phi(\bm{R},\bm{r}) \underline{\underline{\Gamma}}_+^{-1} + \frac{1}{\gamma} \frac{2}{1+\kappa^2} \bm{F} \bm{\sigma}_1 \right]=\\
    &=-\frac{1}{8D_R}\int d \bm{r}\, v_a\left(\bm{R} + \frac{1}{2}\underline{\underline{\Gamma}}_- \bm{r}\right) \nabla_{\bm{R}} \left[ v_a\left(\bm{R} + \frac{1}{2}\underline{\underline{\Gamma}}_- \bm{r}\right) \phi(\bm{R},\bm{r}) \right]+\\
    &+\frac{1}{8D_R}\int d \bm{r}\, \left(\nabla_{\bm{R}} \cdot \underline{\underline{\Gamma}}_- v_a\left(\bm{R} + \frac{1}{2}\underline{\underline{\Gamma}}_- \bm{r}\right)\right)  \cdot \left[ v_a\left(\bm{R} + \frac{1}{2}\underline{\underline{\Gamma}}_- \bm{r}\right) \phi(\bm{R},\bm{r}) \underline{\underline{\Gamma}}_+^{-1} + \frac{1}{\gamma} \frac{2}{1+\kappa^2} \bm{F} \bm{\sigma}_1 \right]
\end{split}
\label{Jsigma1_dimerB}
\end{equation}
where in the last step we used the relation:
\begin{equation}
    \nabla_{\bm{r}}v_a\left(\bm{R} + \frac{1}{2}\underline{\underline{\Gamma}}_- \bm{r}\right)=\frac{1}{2} \nabla_{\bm{R}} \cdot \left[ \underline{\underline{\Gamma}}_- v_a\left(\bm{R} + \frac{1}{2}\underline{\underline{\Gamma}}_- \bm{r}\right) \right]
\end{equation}
Analogously to the first section, we define the integral:
\begin{equation}
    \bm{I}_1=\int d\bm{r}\, \left[ \nabla_{\bm{R}} \cdot \underline{\underline{\Gamma}}_- v_a\left(\bm{R} + \frac{1}{2}\underline{\underline{\Gamma}}_- \bm{r}\right) \right] \cdot \left[ \bm{F} \bm{\sigma}_1\right]
\end{equation}
Using \eqref{sigma1_dimerB} and keeping only the terms contributing with first order gradients we have:
\begin{equation}
    \begin{split}
        \bm{I}_1&=\frac{1}{D_R}\int d\bm{r}\, \left[ \nabla_{\bm{R}} \cdot\left( \underline{\underline{\Gamma}}_- v_a\left(\bm{R} + \frac{1}{2}\underline{\underline{\Gamma}}_- \bm{r}\right) \right) \cdot \bm{F} \right]  \Bigg\{ -\nabla_{\bm{r}} \cdot \left[ v_a\left(\bm{R} + \frac{1}{2}\underline{\underline{\Gamma}}_- \bm{r}\right) \phi(\bm{R},\bm{r}) \underline{\underline{\Gamma}}_+^{-1} + \frac{1}{\gamma} \frac{2}{1+\kappa^2} \bm{F} \bm{\sigma}_1  \right] \Bigg\}=\\
        &=-\frac{k_s}{D_R}\int d\bm{r}\, \left[ \nabla_{\bm{R}} \cdot\left( \underline{\underline{\Gamma}}_- v_a\left(\bm{R} + \frac{1}{2}\underline{\underline{\Gamma}}_- \bm{r}\right) \right) \cdot \underline{\underline{A}}^T \right]   \cdot \left[ v_a\left(\bm{R} + \frac{1}{2}\underline{\underline{\Gamma}}_- \bm{r}\right) \phi(\bm{R},\bm{r}) \underline{\underline{\Gamma}}_+^{-1} + \frac{1}{\gamma} \frac{2}{1+\kappa^2} \bm{F} \bm{\sigma}_1  \right] 
    \end{split}
\end{equation}
where $\nabla_{\bm{r}}\bm{F}=-k_s\underline{\underline{A}}$. In the case of harmonic interaction potential with zero rest length, $\underline{\underline{A}}=\bm{1}$. Thus:
\begin{equation}
\begin{split}
    \bm{I}_1&=-\frac{k_s}{D_R}\int d\bm{r}\, \left[ \nabla_{\bm{R}} \cdot\left( \underline{\underline{\Gamma}}_- v_a\left(\bm{R} + \frac{1}{2}\underline{\underline{\Gamma}}_- \bm{r}\right) \right) \right]   \cdot \left[ v_a\left(\bm{R} + \frac{1}{2}\underline{\underline{\Gamma}}_- \bm{r}\right) \phi(\bm{R},\bm{r}) \underline{\underline{\Gamma}}_+^{-1} + \frac{1}{\gamma} \frac{2}{1+\kappa^2} \bm{F} \bm{\sigma}_1  \right] = \\
    &=-\frac{k_s}{D_R}\frac{1}{\gamma} \frac{2}{1+\kappa^2}\bm{I}_1 - \frac{k_s}{D_R} \int d\bm{r} \, v_a\left(\bm{R} + \frac{1}{2}\underline{\underline{\Gamma}}_- \bm{r}\right)\phi(\bm{R},\bm{r}) \left( \nabla_{\bm{R}} v_a\left(\bm{R} + \frac{1}{2}\underline{\underline{\Gamma}}_- \bm{r}\right) \right) \cdot \underline{\underline{\Gamma}}_- \underline{\underline{\Gamma}}_+^{-1}=\\
    &=-\frac{k_s}{D_R} \frac{1}{1+\frac{k_s}{D_R}\frac{1}{\gamma} \frac{2}{1+\kappa^2}} \int d\bm{r} \, v_a\left(\bm{R} + \frac{1}{2}\underline{\underline{\Gamma}}_- \bm{r}\right)\phi(\bm{R},\bm{r}) \left( \nabla_{\bm{R}} v_a\left(\bm{R} + \frac{1}{2}\underline{\underline{\Gamma}}_- \bm{r}\right) \right) \cdot \underline{\underline{\Gamma}}_- \underline{\underline{\Gamma}}_+^{-1}
    \end{split}
\end{equation}
Plugging this expression back in \eqref{Jsigma1_dimerB}, we obtain:
\begin{equation}
    \begin{split}
    \bm{J}_{\bm{\sigma}_1}=&-\frac{1}{8D_R}\int d \bm{r}\, v_a\left(\bm{R} + \frac{1}{2}\underline{\underline{\Gamma}}_- \bm{r}\right) \nabla_{\bm{R}} \left[ v_a\left(\bm{R} + \frac{1}{2}\underline{\underline{\Gamma}}_- \bm{r}\right) \phi(\bm{R},\bm{r}) \right]+\\
    &+\left[\frac{1}{8D_R} -\frac{1}{4}\frac{k_s}{D_R} \frac{1}{\gamma D_R (1+\kappa^2) + 2k_s}\right]\int d \bm{r}\, v_a\left(\bm{R} + \frac{1}{2}\underline{\underline{\Gamma}}_- \bm{r}\right) \phi(\bm{R},\bm{r}) \left(\nabla_{\bm{R}} v_a\left(\bm{R} + \frac{1}{2}\underline{\underline{\Gamma}}_- \bm{r}\right)\right)  \cdot \underline{\underline{\Gamma}}_-  \underline{\underline{\Gamma}}_+^{-1}
    \end{split}
\end{equation}
An analogous derivation can be done for $\bm{J}_{\bm{\sigma}_2}$, whose expression is:
\begin{equation}
    \begin{split}
    \bm{J}_{\bm{\sigma}_2}=&-\frac{1}{8D_R}\int d \bm{r}\, v_a\left(\bm{R} - \frac{1}{2}\underline{\underline{\Gamma}}_+ \bm{r}\right) \nabla_{\bm{R}} \left[ v_a\left(\bm{R} - \frac{1}{2}\underline{\underline{\Gamma}}_+ \bm{r}\right) \phi(\bm{R},\bm{r}) \right]+\\
    &+\left[\frac{1}{8D_R} -\frac{1}{4}\frac{k_s}{D_R} \frac{1}{\gamma D_R (1+\kappa^2) + 2k_s}\right]\int d \bm{r}\, v_a\left(\bm{R} - \frac{1}{2}\underline{\underline{\Gamma}}_+ \bm{r}\right) \phi(\bm{R},\bm{r}) \left(\nabla_{\bm{R}} v_a\left(\bm{R} - \frac{1}{2}\underline{\underline{\Gamma}}_+ \bm{r}\right)\right)  \cdot \underline{\underline{\Gamma}}_+  \underline{\underline{\Gamma}}_-^{-1}
    \end{split}
\end{equation}
In the regime where the typical distance between the two monomers is much smaller than the activity gradient, we can approximate $\phi(\bm{R},t)\approx \rho(\bm{R},t) \delta(\bm{r})$ inside the integrals (as in Section \ref{dimer_of_ACPs}). This approximation leads to:
\begin{equation}
    \begin{split}
    \bm{J}=\bm{J}_{\bm{\sigma}_1}+\bm{J}_{\bm{\sigma}_2}+\bm{J}_D=&-\frac{1}{4D_R} v_a\left(\bm{R}\right) \nabla_{\bm{R}} \left[ v_a\left(\bm{R}\right) \rho(\bm{R}) \right]- \frac{D_T}{2} \nabla_{\bm{R}} \rho(\bm{R}) + \\
    &+\left[\frac{1}{8D_R} -\frac{1}{4}\frac{k_s}{D_R} \frac{1}{\gamma D_R (1+\kappa^2) + 2k_s}\right]\frac{1}{2}\rho(\bm{R})\nabla_{\bm{R}} v_a(\bm{R})^2 \cdot \left[\underline{\underline{\Gamma}}_-  \underline{\underline{\Gamma}}_+^{-1} + \underline{\underline{\Gamma}}_+  \underline{\underline{\Gamma}}_-^{-1}\right]
    \end{split}
\end{equation}
This expression can be further simplified being
\begin{equation}
    \underline{\underline{\Gamma}}_-  \underline{\underline{\Gamma}}_+^{-1} + \underline{\underline{\Gamma}}_+  \underline{\underline{\Gamma}}_-^{-1}=2\frac{1-\kappa^2}{1+\kappa^2} \bm{1}
\end{equation}
Thus:
\begin{equation}
\begin{split}
    \bm{J}=&\left[ -\frac{1}{8D_R} + \left( \frac{1}{8D_R} -\frac{1}{4}\frac{k_s}{D_R} \frac{1}{\gamma D_R (1+\kappa^2) + 2k_s} \right) \frac{1-\kappa^2}{1+\kappa^2}\right] \left( \nabla_{\bm{R}} v_a(\bm{R})^2\right) \rho(\bm{R})+\\&-\left[\frac{1}{4D_R}v_a(\bm{R})^2 + \frac{D_T}{2} \right] \nabla_{\bm{R}}\rho(\bm{R})
    \end{split}
\end{equation}
Therefore, \eqref{coarse_grained_fpe_dimerB} can be rewritten as the effective coarse grained FPE
\begin{equation}
    \begin{split}
        &\partial_t \rho(\bm{R})=-\nabla_{\bm{R}} \cdot \left[ \bm{V}(\bm{R}) \rho(\bm{R})-D(\bm{R})\nabla_{\bm{R}}\rho(\bm{R}) \right]\\
        \\
        &\bm{V}(\bm{R})=\left[ -\frac{1}{8D_R} + \left( \frac{1}{8D_R} -\frac{1}{4}\frac{k_s}{D_R} \frac{1}{\gamma D_R (1+\kappa^2) + 2k_s} \right) \frac{1-\kappa^2}{1+\kappa^2}\right] \left( \nabla_{\bm{R}} v_a(\bm{R})^2\right)\\
       \\ &D(\bm{R})=\frac{1}{4D_R}v_a(\bm{R})^2 + \frac{D_T}{2} 
    \end{split}
\end{equation}
Being $\bm{V}(\bm{R})=-\frac{\epsilon}{2}\nabla_{R}D(\bm{R})$ with
\begin{equation}
    \epsilon=1-\left(1-\frac{2k_s}{\gamma D_R (1+\kappa^2) + 2k_s} \right)\frac{1-\kappa^2}{1+\kappa^2}
\end{equation}
we can easily derive the steady state density
\begin{equation}
    \rho(\bm{R})\propto \left[ \frac{1}{4D_R}v_a(\bm{R})^2 + \frac{D_T}{2}  \right]^{-\frac{\epsilon}{2}}
\end{equation}

\end{document}